\documentclass[prc,twocolumn,epsfig,nofootinbib,floatfix,showpacs]{revtex4-2}
\usepackage{graphics}
\usepackage{epsfig}
\usepackage{amsfonts}
\usepackage{amsmath}
\usepackage{bm}
\usepackage{color}

\begin{document}

\title{Low-energy spectra of nobelium isotopes: Skyrme
random-phase-approximation analysis}

\author{ V.O. Nesterenko $^{1,2}$, M.A. Mardyban $^{1,2}$, A. Repko $^{3}$, R.V. Jolos$^{1,2}$,
P.-G. Reinhard $^{4}$ and Alan A. Dzhioev$^{1}$}
\affiliation{$^1$
Laboratory of Theoretical Physics, Joint Institute for Nuclear Research.
141980, Dubna, Moscow region, Russia}
\affiliation{$^2$
Dubna State University. 141982, Dubna, Moscow region, Russia}
\affiliation{$^3$
Institute of Physics, Slovak Academy of Sciences, 84511 Bratislava, Slovakia}
\affiliation{$^4$
Institute for Theoretical Physics II, University of Erlangen, D-91058, Erlangen, Germany}

\date{\today}

\begin{abstract}

Low-energy spectra in the isotopic chain  $^{250-262}$No are systematically investigated within the
fully self-consistent Quasiparticle Random-Phase-Approximation (QRPA)  using Skyrme forces SLy4,
SLy6, SkM* and SVbas. QRPA states of multipolarity $\lambda\mu$=20, 22, 30, 31, 32, 33, 43, 44 and 98
are considered. The main attention is paid to isotopes $^{252}$No and $^{254}$No where the most
extensive experimental spectroscopic information is available. In these two nuclei, a reasonable
description of $K^{\pi}=8^-, 2^-$and $3^+$  isomers is obtained with forces SLy4 and SLy6.
The disputed  $8^-$ isomer in $^{254}$No is assigned as neutron two-quasiparticle configuration
$nn[734\uparrow,613\uparrow]$. The isomers are additionally analyzed using Skyrme functionals UNEDF1,
UNEDF2 and UNEDF1$^{\rm SO}$. At the energies 1.2 - 1.4 MeV, the 2qp $K$-isomers $4^-, 7^-$ in $^{252}$No
and $4^-, 6^-, 7^-$ in $^{254}$No are also predicted. In $^{254}$No, the $K^{\pi}=3^+$ isomer should
be accompanied by the nearby $K^{\pi}=4^+$ counterpart.
It is shown that, in the chain  $^{250-262}$No, some features of $^{252}$No and $^{254}$No should exhibit essential
irregularities caused by a noticeable shell gap in the neutron single-particle spectrum and corresponding reduction
of the neutron pairing. In particular, low-energy pairing vibrational $K^{\pi}=0^+$ states in $^{252,254}$No
are predicted.
\end{abstract}
\maketitle

\section{Introduction}

Spectroscopy of transfermium nuclei (Pu, Cm, Cf, Fm, No) is presently of a large interest,
see reviews~\cite{Herz08,Herz23,Ack24} and references therein. These nuclei provide
valuable information on the single-particle spectra and pairing in heavy nuclei,
which in turn can be useful for understanding the features  of neighbouring
superheavy nuclei. Though experimental spectroscopic information for transfermium
nuclei is permanently increasing (see early~\cite{Herz_Nature2006,Tan_PRL2006,Rob_PRC2008,Hess_EPJA10,Clark_PLB10}
 and  recent~\cite{Tez22,Forge_JPCS2023,Forge_PhD,Wahid_PRC25} experimental studies),
it still remains rather scarce. Among transfermium nuclei, nobelium isotopes have, perhaps,
the most extensive experimental spectroscopic data  \cite{Herz08,Herz23,nndc}.

In nobelium isotopes, $^{252}$No and $^{254}$No are most investigated~\cite{Herz08,Herz23}.
In these two nuclei, long ground-state rotational bands (up to $I^{\pi}=20^+$ in $^{252}$No
and $I^{\pi}=24^+$ in $^{254}$No) are observed.
Already for two decades, two-quasiparticle (2qp) $K^{\pi}=8^-$ isomers (at
1.254 MeV in $^{252}$No and 1.295 MeV in $^{254}$No) are scrutinized
by both experiment and theory. Some low-energy rotational bands with band heads
$K^{\pi}=2^-$ at 0.929 MeV in $^{252}$No and $K^{\pi}=3^+$ at 0.987 MeV in $^{254}$No
are found. Quite recently nonrotational states $K^{\pi}=4^+$ (1.203 MeV) and $0^+$ (0.888 MeV)
in $^{254}$No were tentatively observed~\cite{Forge_JPCS2023}. The $0^+$ state was
treated as a result of the coupling between "normal" deformed and  superdeformed shapes.

While the experimental information on nobelium isotopes is generally poor,
there are already a lot of theoretical studies and predictions. The quadrupole
moments and corresponding quadrupole deformations were estimated, see
e.g.~\cite{Bar_PRC2005,Kle_PRC2008,De06,Afan_PRC2003,Afan_PRC2013,Dob_NPA2015}.
Single-particle (s-p) spectra, moments of inertia and pairing effects were
systematically explored within the cranked relativistic Hartree-Bogoliubov (CRHB) theory
\cite{Afan_PRC2003,Afan_PRC2013}. The results from different  density functional
theories (DFT), including relativistic covariant~\cite{Vre_PR2005}, Skyrme~\cite{Ben_RMP2003} and
Gogny~\cite{Gogny}, were critically compared~\cite{Dob_NPA2015}.
Various multipole low-energy two-quasiparticle
(2qp) excitations were considered in self-consistent calculations with Gogny force~\cite{De06}.
The investigation of s-p spectra, deformations, pairing and $\alpha$-decay features
in nobelium isotopes was recently performed within the symmetry-unrestricted nonrelativistic mean-field
model employing a shell correction analysis~\cite{Guo_PRC24}.

Further, long rotational bands in $^{252,254}$No were investigated within
CRHB theory~\cite{Afan_PRC2003,Afan_PRC2013,Xu_PRL2024},
cranked shell model \cite{Zhang_PRC2012,DN22,DN24}, Skyrme-Hartree-Fock-Bogoliubov and
Lipkin-Nogami (LN) methods \cite{Shi_PRC2014}, and interacting-boson model
(IBM)~\cite{Ef-Iz_PAN2021}.

The relation between the lowest $1^-$ states  and reflection asymmetry in nobelium isotopes was
studied~\cite{Mardyban}. Within the cluster model, the alternative parity bands
based on the ground $0^+$ and vibrational $1^-$ states were predicted as a consequence of the
reflection asymmetry~\cite{Shnei_PRC2006}.
The scissor mode was measured \cite{Oslo_EPJA22} and analyzed within
the Wigner Function Method~\cite{Balb_EPJA24}.

$K$-isomers in $^{252,254}$No were also widely explored, see
e.g.~\cite{DN22,DN24,Sol91,Forge_JPCS2023,Forge_PhD,Wahid_PRC25,Herz_Nature2006,Tan_PRL2006,
Rob_PRC2008,Ada_PRC2010,Jol11,He_CPC2020,Zhang_PRC25}. Two-quasiparticle
$K$-isomers $8^-, 2^-$ and $3^+$  were considered in the framework of
schematic QRPA approach~\cite{Sol91,Jol11} using Woods-Saxon (WS) potential and
pairing blocking at the Bardeen-Cooper-Schrieffer (BCS) level.
In the two-quasiparticle approximation, the same isomers were analyzed with
the universal WS potential and pairing treated within LN
procedure~\cite{Tan_PRL2006,Rob_PRC2008,Wahid_PRC25}.  In the studies
~\cite{Tan_PRL2006,Rob_PRC2008,Clark_PLB10,Wahid_PRC25}, the complete decay
scheme for isomers $8^-$ was suggested (though being
disputed in the recent experiment~\cite{Forge_JPCS2023,Forge_PhD}).
Within the cranking shell model (Nilsson potential with hexadecapole deformation) + pairing
with the particle-number projection (CSM+PNP), a nice description of the energy of $8^-$-isomer
in $^{254}$No was obtained \cite{He_CPC2020}.
Using the two-center shell model,
$\alpha$-decays relevant for the $K$-isomers were  explored~\cite{Ada_PRC2010}.
In recent shell model calculations~\cite{DN24}, the energies of $8^-$ and $3^+$ isomers
in $^{254}$No were well reproduced. Nevertheless, despite impressive theoretical progress,
there are still some open problems. For example, the 2qp assignment of $8^-$ isomer in
$^{254}$No is still disputed~\cite{Herz08,Herz23,Forge_JPCS2023,Forge_PhD,Wahid_PRC25}.

A significant part of the previous theoretical studies was performed within non self-consistent
models. However, even modern self-consistent models still produce
rather different results and have troubles with the description of shell structures and other
features observed in experiments, see discussion \cite{Herz08,Herz23,Dob_NPA2015}.
Moreover, most of the previous studies were devoted to description of
the mean-field, pairing, rotation  and decay properties.
At the same time, the self-consistent predictions for low-energy  non-rotational states still
remain very limited.

In this connection, we present here a thorough investigation of non-rotational
low-energy spectra in $^{250-262}$No within fully self-consistent QRPA
model~\cite{QRPA,Rep_EPJA2017_pairing,Kva_EPJA2019_spurious}
with Skyrme forces SLy4~\cite{SLy6},  SLy6~\cite{SLy6},
SkM*~\cite{SkM} and SVbas~\cite{SVbas}. The main attention is paid to
$^{252}$No and $^{254}$No, where we consider the observed $K$-isomers (as relevant test cases)
and the lowest one-phonon states with
$K^{\pi}=0^+,2^+,3^+,4^+,0^-,1^-,2^-$ up to excitation energy 2-2.5 MeV.
Note that at N=152 a shell gap in the neutron  single-particle spectrum was
predicted~\cite{Rob_PRC2008}. So one may expect in $^{254}$No and nearby nuclei a
significant reduction of the neutron pairing  and subsequent irregularities in features.
To test the pairing in these nuclei, we calculate the lowest pairing vibrational one-phonon
$K^{\pi}=0^+$ states which are a good indicator of the actual pairing.

It is important that the applied Skyrme forces have different isoscalar effective masses
$m^*/m$ affecting the distribution of single-particle states near the Fermi level~\cite{Ne_PRC2016}.
Moreover, the applied Skyrme forces exploit different kinds of the pairing: volume in SLy4, SLy6
and SkM* and surface density-dependent in SVbas. The SLy4 and SLy6 belong to one family of Skyrme forces
\cite{SLy6} and are rather similar. We use both them to demonstrate that in heavy nuclei even similar forces
can lead to essentially different BCS description of the pairing.

Special attention is paid to isomers $8^-$ in $^{252,254}$No. Since in these isotopes
the drop of the neutron pairing can take place, we additionally scrutinize this case with
Skyrme functionals UNEDF1~\cite{Kor_UNEDF1}, UNEDF2~\cite{Kor_UNEDF2} and UNEDF1$^{SO}$~\cite{Shi_PRC2014}
employing LN procedure for the pairing.

The paper is organized as follows. In Sec. II, the model and details of the calculations are
briefly described. In Sec. III, the trends and irregularities in features of $^{250-262}$No are discussed.
The single-particle spectra in  $^{252,254}$No are demonstrated. The pairing features and two-neutron
mass staggering are considered  as fingerprints of shell gaps. In Sec. IV, various multipole low-energy 2qp
and one-phonon QRPA states in $^{252,254}$No are analyzed, $K$-isomers are thoroughly discussed,
pairing vibrational states are demonstrated. In Sec. V, the conclusions  are drawn. In Appendix A,
 the UNEDF1, UNEDF2 and UNEDF1$^{\rm SO}$ results for isomers in $^{252,254}$No are presented.

\section{Model and calculation details}

The calculations are performed within QRPA model
\cite{QRPA,Rep_EPJA2017_pairing,Kva_EPJA2019_spurious}
 based on Skyrme functional \cite{Ben_RMP2003}. The model is fully self-consistent since
 i) both mean field and residual interaction are derived from the same Skyrme functional,
 ii) the contributions of all time-even densities and time-odd currents from the
 functional are taken  into account, iii) both particle-hole and pairing-induced
 particle-particle channels are included, iv) the Coulomb (direct and exchange) parts
 are involved in both mean field and residual interaction.

\begin{table}[t]  
\caption{Isoscalar effective mass $m^*/m$, proton and neutron pairing constants
$G_p$ and $G_n$, kind of pairing (volume, surface) and IS and IV spin-orbit
parameters $b_4$ and $b'_4$ for SLy4, SLy6, SkM* and SVbas. The surface pairing
parameter is $\rho_{\rm pair}$=0.2011 ${\rm fm}^{-3}$.}
\begin{tabular}{|c|c|c|c|c|c|c|c|}
\hline
force   & $m^*/m$ &  kind of  & $G_p$        & $G_n$  & $b_4$ & $b'_4$\\
\cline{4-7}
        & & pairing & \multicolumn{2}{|c|}{(MeV fm$^3$)}  & \multicolumn{2}{|c|}{(MeV fm$^5$)}\\
\hline
SLy4   & 0.70 & volume & 295.37 &  286.67 &  61.5 & 61.5 \\
SLy6   & 0.69 & volume & 298.76 &  288.52  &  61.0 & 61.0 \\
SkM*   & 0.79  & volume & 279.08 & 258.96 &  65.0 & 65.0 \\
SVbas & 0.90   & surface & 674.62 & 606.90 &  62.3 & 34.1 \\
\hline
\end{tabular}
\label{mG}
\end{table}

 A representative set of Skyrme parametrizations (SLy4~\cite{SLy6}, SLy6 \cite{SLy6}, SkM* \cite{SkM} and
 SVbas~\cite{SVbas}) with various isoscalar effective masses $m^*/m$ and different kinds of the
 paring  is used, see Table~\ref{mG}. Besides, these forces have different isoscalar (IS) and isovector
 (IV) spin-orbit parameters. As mentioned above, SLy4 and SLy6 belong to the same family of Skyrme forces
 and are rather similar. As compared with the basic parametrization SLy4, the parameters of SLy6
 were fitted  with an additional two-body center-of-mass correction in the energy functional.

 The method QRPA is implemented in a matrix form \cite{QRPA}. Spurious
 admixtures caused  by  violation of the translational and rotational invariance and by pairing-induced
mixed particle number are removed using the technique of Ref.~\cite{Kva_EPJA2019_spurious}. The states
 $K^{\pi}=0^+,2^+,3^+,4^+,0^-,1^-,2^-,3^-,8^-$ are calculated as one-phonon excitations of multipolarity
 $\lambda\mu=$20,22,43,44,30,31,32,33,98, respectively.

The single-particle spectra and pairing characteristics are calculated with the
code SKYAX \cite{Skyax} using two-dimensional (2D) grid in cylindrical coordinates.
The grid step 0.7 fm and calculation box up to 3 nuclear radii are employed. All
proton and neutron s-p levels from the bottom of the potential well
up to +40 MeV are taken into account. For example, SLy6 calculations  for $^{254}$No
employ 1779 proton and 2035 neutron s-p levels.

The equilibrium axial quadrupole deformations
\begin{equation}
\beta=\frac{4\pi}{3} \frac{Q_{2}}{AR^2}
\label{beta}
\end{equation}
and the related intrinsic quadrupole moments
\begin{equation}
Q_{2}=\sum_{q=p,n}\int d^3r (2z^2-x^2-y^2)\rho_q(\textbf{r})
\label{Q2}
\end{equation}
are obtained at the minimum of the total nuclear energy. Here, $A$ is the mass number,
 R=r$_0$A$^{1/3}$ with  r$_0$ =1.2 fm is the nuclear radius and $\rho_q(\textbf{r})$
 are proton ($q=p$) and neutron ($q=n$) densities in the ground state (g.s.).

For the pairing,
the zero-range pairing interaction~\cite{Be00}
\begin{equation}
V_q^{\rm pair} (\textbf{r}, \textbf{r}') = - G_q \left[ 1-\eta \left(\frac{\rho(\textbf{r})}{\rho_{\rm pair}} \right) \right]\delta(\textbf{r}-\textbf{r}') ,
\label{pairing}
\end{equation}
is used, where $G_q$ are proton and neutron pairing strength constants (shown in Table~\ref{mG}).
The pairing is treated within the BCS scheme~\cite{Rep_EPJA2017_pairing}.
The strength constants are fitted to reproduce empirical pairing
gaps in selected isotopic and isotonic chains~\cite{Rein_pairing}.
To cope with a divergent character of zero-range pairing forces,
the energy-dependent cut-off is used \cite{Rep_EPJA2017_pairing,Be00}.
The volume density-independent and  surface density-dependent
cases of pairing are realized at $\eta$=0 and 1, respectively.
The model parameter $\rho_{\rm pair}$ is 0.2011 ${\rm fm}^{-3}$ is determined
in the SVbas fit~\cite{SVbas}.

The pairing (\ref{pairing}) results in state-dependent pairing gaps $\Delta_q(i)$ \cite{SVbas,Be00}.
Then it is convenient to evaluate the pairing by  {\it average} spectral pairing gaps
\begin{equation}
 \Delta_q = \frac{\sum_{i \in q} v_i u_i \Delta_q(i)}{\sum_{i \in q} v_i u_i}
\end{equation}
where $v_i$ and $u_i$ are Bogoliubov pairing factors for quasiparticle states $i$.
These spectral gaps usually well correspond to the five-point experimental
gaps $\Delta_q^{(5)}$ in mid-shell regions~\cite{Be00}.

The calculations use a large 2qp configuration space. The energy weighted sum rules
(EWSR) for $\lambda$=0,1,2,3 are exhausted by 90-100$\%$.

The moments of inertia (MoI) are calculated using the Thouless-Valatin expression~\cite{TV}
\begin{equation}
J=2\sum_\nu\frac{|\langle \nu|I_x|0 \rangle|^2}{E_\nu}
\end{equation}
where $I_x$ is x-component of the operator of the total angular moment $I$,
$|0\rangle$ is the QRPA vacuum and $|\nu\rangle$ is the excited QRPA state with the energy E$_\nu$.
For rotational energies, we use formula \cite{BM2}
\begin{equation}
E_I=\frac{I(I+1)}{2J} .
\label{E_I}
\end{equation}

The reduced probabilities ($\lambda > 1)$
\begin{equation}
   B(E\lambda\mu)= e^2 (2-\delta_{\mu,0})|\langle\nu |F(E\lambda\mu)|0\rangle|^2
\end{equation}
for $E\lambda\mu$ -transitions from the ground state $|0\rangle$ with
$I^{\pi}K=0^+0_{\rm{gs}}$  to the QRPA state $|\nu\rangle$
with $I^{\pi}K$ ($I=\lambda, K=\mu, \pi=(-1)^{\lambda}$)
are computed with the transition operator
\begin{equation}
   F(E\lambda\mu)=\sum_{i=1}^{Z} [r^{\lambda} Y_{\lambda\mu}(\Omega)]_i .
\label{EL_operator}
\end{equation}
The transition probabilities are estimated in
Weisskopf units W.u.$=[3 R^{\lambda}/(\lambda +3)]^2/(4\pi) \; e^2 \rm{fm}^{2\lambda}$.

The normalized  $E0$ transition probability  for the decay
$0^+0_{\nu} \to 0^+0_{\rm{gs}}$ is described by
the dimensionless value
\begin{equation}
   \rho^2(E0)= \frac{1}{R^4}|\langle 0 |F(E0)|\nu \rangle|^2 \; \text{with} \; F(E0)=\sum_{i=1}^{Z} r^2_i.
\end{equation}

\begin{table}[h] 
\caption{The calculated and experimental~\cite{nndc} quadrupole deformations $\beta$,
moments of inertia $J$, proton $\Delta_p$ and neutron  $\Delta_n$
pairing spectral average gaps,  energies $E(2^+_1)$ of the lowest rotational
$I^{\pi}=2^+$  states in $^{252,254}$No.}
\begin{tabular}{|c|c|c|c|c|c|c|c|}
\hline
Nucleus      &                               & SLy4  & SLy6  & SkM* & SVbas & exp \\
\hline
$^{252}$No            & $\beta$               & 0.303 &   0.300    &   0.306   &   0.299  & \\
                      &    $J$ ($\hbar^2$/MeV)& 79.9 &   76.8    &   74.1   &   58.0 &  64.7\\
                      &    $\Delta_p$ (MeV)   & 0.48 &    0.52   &   0.38   &   0.56 & 0.67 \\
                      &    $\Delta_n$ (MeV)   & 0.32 &    0.33   &   0.51   &  0.70  &  0.71 \\
 	                  &     $E(2^+_1)$ (keV)  &  38  &      39   &      40   &    52 &   46\\
\hline
$^{254}$No       &     $\beta$                & 0.304 &    0.298    &  0.304    &   0.298 &  \\	
 	             &    $J$ ($\hbar^2$/MeV)     & 76.0  & 78.3    &   80.8   &   59.3 & 67.9 \\
 	             &     $\Delta_p$ (MeV)       &  0.47 &    0.52    &   0.40   &   0.55 & 0.66\\
	             &     $\Delta_n$ (MeV)]      &  0.28 &  0.05   &   0.36  &  0.67  &  0.71 \\
 	             &     $E(2^+_1)$ (keV)       &  39   & 38   &      37   &     50 &   44\\          	             	
\hline
\end{tabular}
\label{No}
\end{table}

\begin{figure}[h] 
\centering
\includegraphics[width=0.44\textwidth]{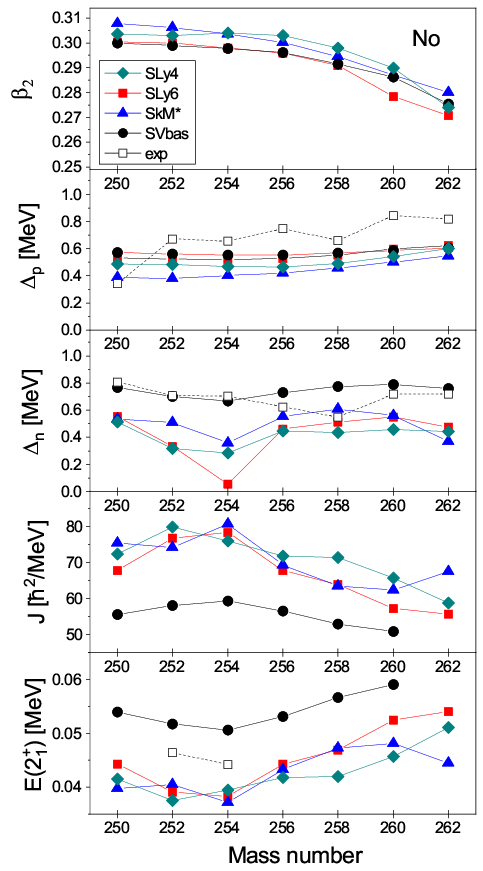}
\vspace{1mm}
\caption{Deformations $\beta$, moments of inertia $J$, spectral average pairing gaps
$\Delta_{n,p}$, and energies $E(2^+_1)$ in $^{250-262}$No, calculated with the forces
SLy4 (green diamonds), SLy6 (red filled squares), SkM* (blue filled
triangles) and SVbas (black filled circles). The experimental values for pairing gaps and
energies $E(2^+_1)$ \cite{nndc} are shown by open squares.}
\label{fig1:No_features}
\end{figure}

 \section{Mean-field and pairing features of nobelium isotopes}

Some calculated  and experimental characteristics of $^{252,254}$No are shown in
Table~\ref{No}. The experimental pairing gaps $\Delta_q$ are obtained by
five-point pairing formula using evaluations for binding energies~\cite{nucl_masses}.
The experimental MoI are evaluated as $J=3/E(2^+_1)$
where $E(2^+_1)$ is the measured energy of $I^{\pi}=2^+$ state of the g.s. rotational band.

Following Table~\ref{No}, the applied forces provide a  rough description
of the experimental $E(2^+_1)$-energies and MoI. To the best of our knowledge, there is
no experimental data for quadrupole
 deformations $\beta$ in $^{252,254}$No.
However, all the forces give similar results for $\beta$, which are
in a good agreement with the values of macroscopic-microscopic model~\cite{Bar_PRC2005},
see the comparison in Ref.~\cite{Kle_PRC2008}. Note that, following the study~\cite{Wang12},
the quadrupole triaxial and octupole deformations in $^{252,254}$No  near the main
minimum of potential energy surface (PES) at $\beta\approx $0.3 are negligible.

\begin{figure}[h] 
\centering
\includegraphics[width=0.48\textwidth]{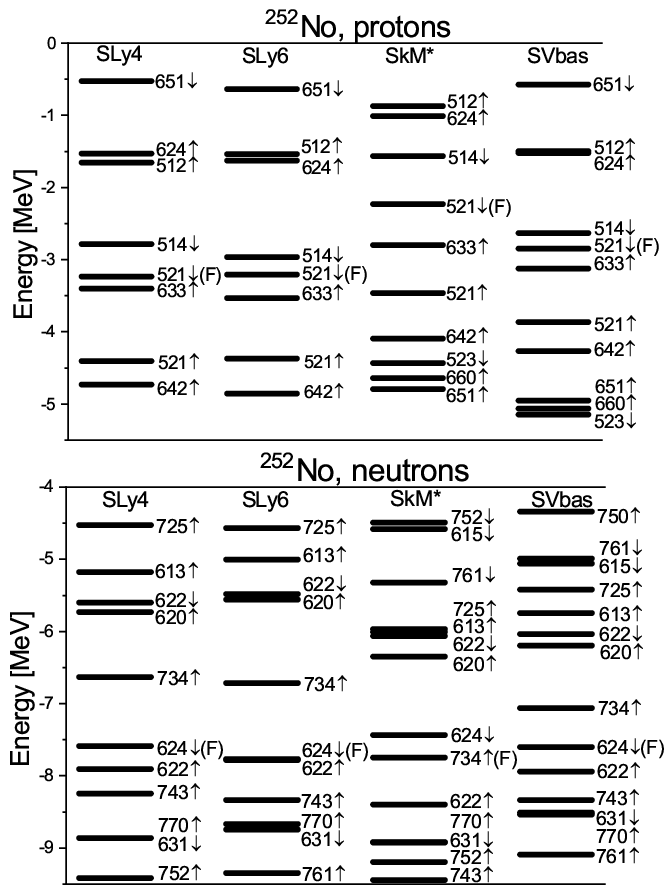}
\caption{SLy4, SLy6, SkM* and SVbas proton and neutron s-p spectra near the Fermi
levels (F) for $^{252}$No.}
\label{fig2:252No_sps}
\end{figure}
\begin{figure}[h] 
\centering
\includegraphics[width=0.48\textwidth]{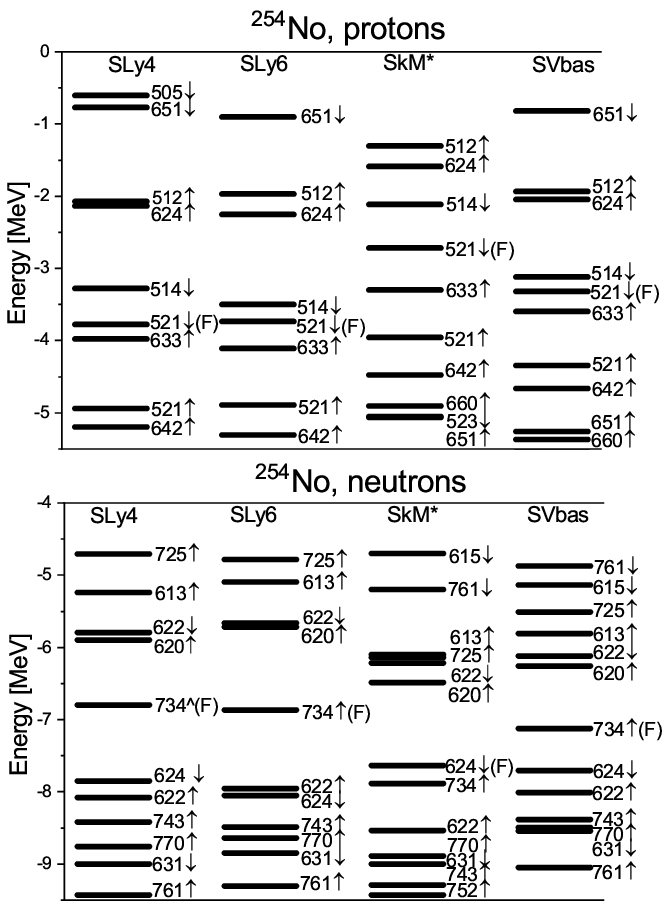}
\caption{The same as in Fig.~\ref{fig2:252No_sps} but for $^{254}$No.}
\label{fig3:254No_sps}
\end{figure}
\begin{figure}[h] 
\centering
\includegraphics[width=0.48\textwidth]{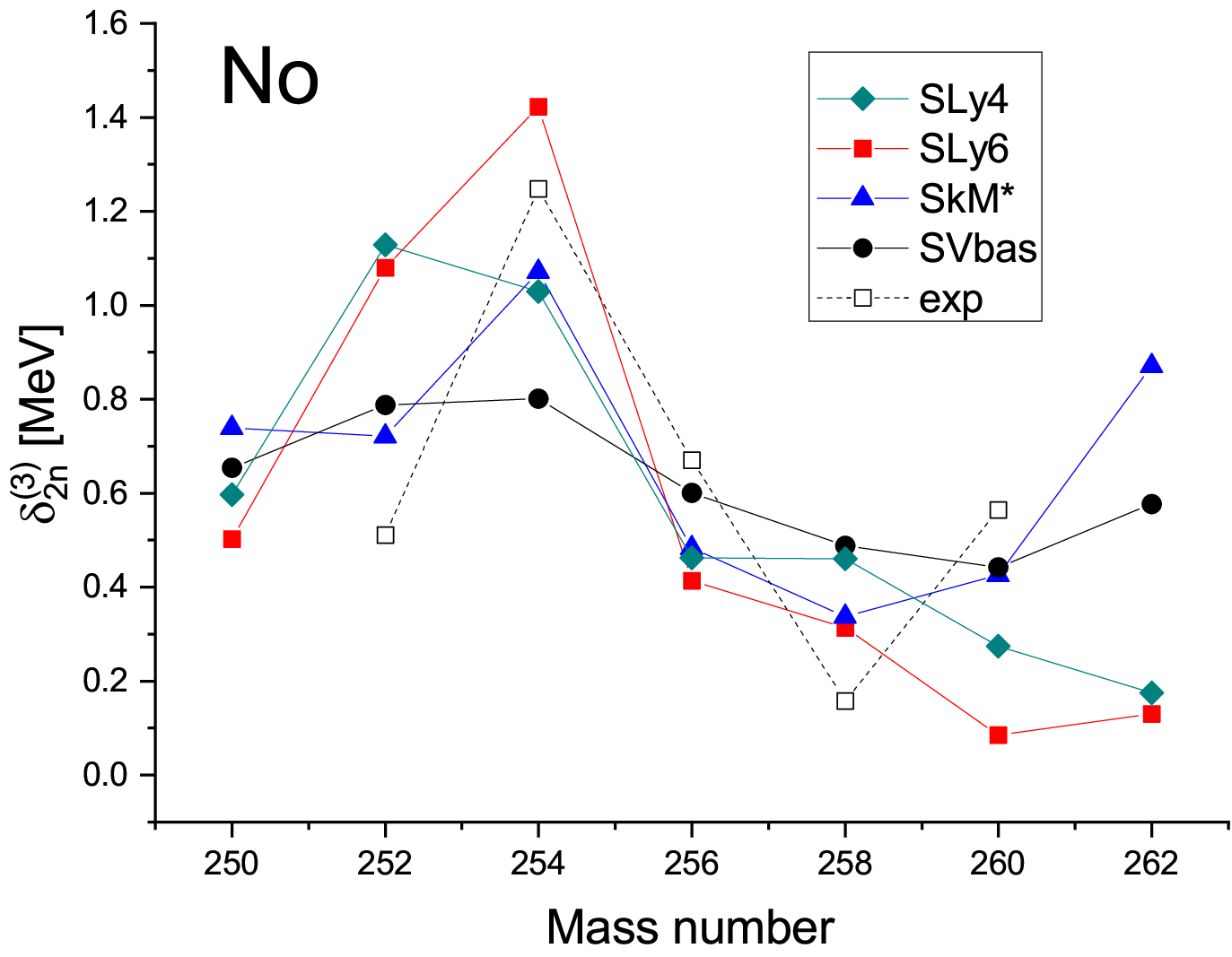}
\caption{SLy4, SLy6, SkM*, SVbas and experimental neutron values $\delta^{(3)}_{2n}$
in nobelium isotopes.}
\label{fig4: S2n}
\end{figure}

Table~\ref{No} shows that SVbas provides the most reasonable description of
the experimental pairing gaps $\Delta_{p,n}$. In particular, SVbas gives
$\Delta_{n} > \Delta_{p}$ in accordance with the experiment. Instead,
SLy4, SLy6 and SkM* demonstrate a weaker neutron pairing and even its collapse in $^{254}$No
for SLy6. The latter is caused by using BCS technique in the case of a weak pairing.
Following our analysis, the similar forces SLy4 and SLy6 give essentially different BCS
results for neutron pairing in $^{254}$No because for SLy4 the density of neutron s-p states
near the Fermi level in this nucleus is slightly larger than for SLy6 (see Fig. 3 for the comparison).
This demonstrates  how fragile could be the calculation results in heavy nuclei.

The collapse of the neutron pairing for SLy6 in $^{254}$No is seen in
Fig.~\ref{fig1:No_features}, where various characteristics of nobelium isotopes are displayed.
SLy4, SkM* and SVbas also give some minima in $\Delta_{n}$
for $^{254}$No but these minima are much softer, especially for SVbas.
Note that a similar distinctive minimum in $\Delta_n$ was also obtained at
$N$=149-151 using Skyrme, Gogny and  covariant relativistic density energy
functionals~\cite{Dob_NPA2015}. Following Fig.~\ref{fig1:No_features},
just the drop of the neutron pairing causes the irregularity in $A$-dependence of MoI and
E$(2^+_1)$. Instead, the deformation $\beta$ in nobelium isotopes changes
smoothly and can lead only to a gradual change of these values.

Fig.~\ref{fig1:No_features} also shows experimental $\Delta_{n,p}$ evaluated by
five-point formula with the nuclear binding energies  from Ref.~\cite{nucl_masses}.
This formula is commonly used to get smooth pairing gaps among the neighboring nuclei,
see discussion \cite{Be00}. It is seen that we rather well describe proton pairing
gaps $\Delta_{p}$ but not neutron gaps $\Delta_{n}$. Unlike our results, the experimental
$\Delta_{n}$ exhibit a minimum  at A=258 and a plateau at  A=252-254.

To understand the origin of $\Delta_{n}$-minimum  in our calculations, let's  consider
SLy4, SLy6, SkM* and SVbas
s-p proton and neutron spectra in $^{252,254}$No, exhibited in Figs.~\ref{fig2:252No_sps} and
\ref{fig3:254No_sps}. The s-p levels in the figures are marked  by Nilsson asymptotic numbers
$[N,n_z,\Lambda]$, where $N$ is the principle quantum number, $n_z$ is the fraction of $N$ along
z-axis and  $\Lambda$ is projection of the orbital moment onto z-axis ~\cite{nilsson}.

Figure~\ref{fig2:252No_sps} for $^{252}$No shows that, for all four Skyrme forces, the proton
Fermi level lies in rather dense spectrum and so we have a significant proton pairing. Instead,
in the neutron case, SLy4 and SLy6 Fermi level  $[624\downarrow]$ lies at the boundary of a wide shell gap,
which leads to a small $\Delta_{n}$=0.32-0.33 MeV. Actually, this particular
neutron spectral configuration is not a true shell gap since it separated by the orbital
$[734\uparrow]$ into two parts. It is rather a region
of a low level density. Nonetheless, for simplicity the term "shell gap" is used hereafter.

In Fig.~\ref{fig3:254No_sps} for $^{254}$No, the SkM* neutron Fermi level
lies near the shell gap and so we have a significant minimum in $\Delta_{n}$ depicted for this force in
Fig.~\ref{fig1:No_features}. For SLy4 and SLy6, the neutron Fermi level $[734\uparrow]$ enters
the center of the shell gap, which results in a small $\Delta_{n}$=0.28 MeV for SLy4 and
collapse of the neutron pairing for SLy6. The difference in SLy4 and SLy6 results is caused
by a different energy gap between Fermi (F) and next (F+1) levels, which is 0.90 MeV in SLy4 and
1.15 MeV in SLy6. The smaller gap for SLy4 prevents the BCS collapse of the neutron pairing.
Our SLy4 and SLy6 results for the shell gap in neutron s-p spectrum in $^{254}$No
agree well with the previous calculations using the Woods-Saxon mean field~\cite{Rob_PRC2008} and
various self-consistent s-p potentials (Skyrme SLy4 and UNEDF2, Gogny DS1 and DM1 and relativistic
NL3*)~\cite{Dob_NPA2015}. These calculations also
 give at $\beta\approx$0.3 a significant neutron shell gap with the Fermi level $[734\uparrow]$ in
the middle of the gap. Besides, a gap at $N$=152 was demonstrated in cranked-shell-model
calculations~\cite{Zhang_PRC25}. In Figs.~\ref{fig2:252No_sps} and \ref{fig3:254No_sps},
SVbas gives the same neutron Fermi levels as SLy4 and SLy6.
Instead, in SkM*, the neutron levels $[734\uparrow]$  and $[624\downarrow]$ are swapped and so
SkM* suggests other neutron Fermi levels in $^{252,254}$No.

A considerable shell gap in the neutron s-p spectra should affect features
of several isotopes (from  $^{250}$No to $^{258}$No) since their Fermi levels lie either near
($[624\downarrow]$, $[622\uparrow]$,  $[620\uparrow]$, $[622\downarrow]$
or inside the gap ($[734\uparrow]$). In particular, one may expect a suppression of $\Delta_{n}$ in
these four isotopes. Then the difference in the minima of calculated and experimental
$\Delta_{n}$, exhibited in Fig.~\ref{fig1:No_features}, is not surprising. Note also that,
in the case of shell gaps, values $\Delta_{q}$ manifest rather mean-field effects than
the paring impact (see a general discussion \cite{Be00}).

\begin{table*}
\caption{Experimental~\cite{nndc} and calculated ground and first excited non-rotational
$I^{\pi}$ states in $Z$-odd and $N$-odd neighbours of $^{252}$No and $^{254}$No.
The experimental assignments in the parentheses are tentative.}
\begin{tabular}{|c|c|c|c|c|c|c|}
\hline
 Nucl.   & exp & SLy4 & SLy6 & SkM* & SVbas \\
\hline
\multicolumn{6}{|c|}{$^{252}$No} \\
\hline
  $^{251}$No &  (7/2$^+$), (1/2$^+$) & 7/2$^+$, 5/2$^+$  &
  7/2$^+$, 5/2$^+$ & 9/2$^-$, 7/2$^+$ & 7/2$^+$, 5/2$^+$ \\
 \cline{1-6}
 $^{253}$No &  (9/2$^-$), 5/2$^+$ & 9/2$^-$, 1/2$^+$  &
  9/2$^-$, 7/2$^+$ & 7/2$^+$, 9/2$^-$ & 9/2$^-$, 7/2$^+$ \\
 \cline{1-6}
 $^{251}$Md &  7/2$^-$, 1/2$^-$ & 1/2$^-$, 7/2$^+$ &
 1/2$^-$, 7/2$^-$ & 1/2$^-$, 7/2$^-$ & 1/2$^-$, 7/2$^-$ \\
\cline{1-6}
 $^{253}$Lr &  (7/2$^-$), (1/2$^-$) & 7/2$^-$, 1/2$^-$ &
 7/2$^-$, 1/2$^-$ & 7/2$^-$, 9/2$^+$ & 7/2$^-$, 1/2$^-$ \\
\hline
\multicolumn{6}{|c|}{$^{254}$No} \\
\hline
  $^{253}$No &  (9/2$^-$), 5/2$^+$ & 9/2$^-$, 1/2$^+$ &
  9/2$^-$, 5/2$^+$ & 7/2$^+$, 9/2$^-$ & 9/2$^-$, 7/2$^+$ \\
  \cline{1-6}
  $^{255}$No &  (1/2$^+)$  & 1/2$^+$, 3/2$^+$ &
   1/2$^+$, 3/2$^+$ & 1/2$^+$, 3/2$^+$ & 1/2$^+$, 3/2$^+$\\
 \cline{1-6}
 $^{253}$Md &  (7/2$^-$) &  1/2$^-$, 7/2$^+$  &
 1/2$^-$, 7/2$^-$ & 1/2$^-$, 7/2$^+$ & 1/2$^-$, 7/2$^-$ \\
\cline{1-6}
  $^{255}$Lr &  (1/2$^-$), (7/2$^-$) & 7/2$^-$, 1/2$^-$ &
  7/2$^-$, 1/2$^-$ & 7/2$^-$, 9/2$^+$ & 7/2$^-$, 1/2$^-$ \\
\hline
\end{tabular}
\label{table3:odd}
\end{table*}

To clarify more the point with the neutron shell gap, it is worth to consider
the neutron mass staggering quantity
\begin{eqnarray}
\label{delta_2n}
\delta^{(3)}_{2n}(N,Z)&=&S_{2n}(N,Z) - S_{2n}(N+2,Z)
 \\
 &=&2B(N,Z)-B(N-2,Z)-B(N+2,Z) ,
 \nonumber
\end{eqnarray}
which is typically used to identify two-neutron shell gaps, see
e.g.~\cite{Dob_NPA2015,Guo_PRC24}. Here, $S_{2n}(N,Z)$  is the two-neutron separation energy
and $B(N,Z)$ is the nuclear binding energy. In Fig.\ref{fig4: S2n}, the calculated
$\delta^{(3)}_{2n}$ are compared with experimental ones evaluated using the experimental binding energies
\cite{nndc,nucl_masses}. It is seen that, in accordance with our SLy6, SkM* and SVbas calculations,
the experimental $\delta^{(3)}_{2n}$ also exhibits a peak at $A$=254. For SLy4, the value $\delta^{(3)}_{2n}$
grows at $A$=252-254 with a maximum at $A$=252, which contradicts the experimental data.
Altogether, Fig.\ref{fig4: S2n} justifies the existence of the neutron shell gap
and its maximal impact  for $^{254}$No.

A similar result for $\delta^{(3)}_{2n}$ was obtained in calculations~\cite{Dob_NPA2015,Guo_PRC24}.
In Ref.~\cite{Guo_PRC24}, the computed pairing gap $\Delta_{n}$ has no a distinctive
minimum at $N$=152. Perhaps, this is caused by using the LN prescription
which prevents the pairing breakdown. However, the LN procedure
does not yet  ensure a proper description of the paring in transfermium nuclei. For example,
in CRHB studies~\cite{Afan_PRC2003,Afan_PRC2013} an additional manual attenuation of the paring strength
 by 10-12$\%$ was used to get a proper description of the moments of inertia.

A quality of the calculated s-p spectra in $^{252,254}$No can be additionally
checked by identification of ground and first excited states on the neighboring
$N$-odd and $Z$-odd nuclei. In Table~\ref{table3:odd}, we compare our results for odd
nuclei with the experimental data~\cite{nndc}. Note that in many cases (marked by
parenthesis in the table) we deal with a tentative experimental assignment of the states.
In our estimation, the first excited state is determined as the closest
to the Fermi level in the s-p spectrum of the odd nucleus. The states in the upper and
bottom parts of the table are determined separately from s-p spectra of $^{252}$No
and $^{254}$No, exhibited in Figs.~\ref{fig2:252No_sps} and~\ref{fig3:254No_sps}.

Table~\ref{table3:odd} shows that SLy6 demonstrates the best performance. It correctly
indicates the ground states in all $N$-odd nuclei and in $^{253}$Lr.  In other $Z$-odd nuclei,
this force permutates the ground and first excited states. The SLy4 and SVbas descriptions
are somewhat worse.  The performance of SkM*
is unsatisfactory. This force reproduces the ground states only in $^{255}$No and $^{253}$Lr. So,
SLy6 and SVbas s-p spectra near the Fermi level are most robust.
The description of the neutron spectra is better than of proton ones.

\section{Low-energy spectra in $^{252,254}$No}

In this section, we analyze low-energy ($E_{\nu} <$ 2.5 MeV)  states  with
$K^{\pi}=0^+, 2^+, 3^+, 4^+, 0^-, 1^-, 2^-, 3^-, 8^-$   in $^{252,254}$No. Since
the forces SLy4 and  SLy6 are rather similar, we provide SLy4 results only for the states
with available experimental data:  $K^{\pi}=0^+, 3^+, 4^+, 2^-, 8^-$.

\subsection{Isomers $K^{\pi}=8^-$}

As a first step, we consider two-quasiparticle $8^-$ isomers in $^{252}$No
and $^{254}$No, discussed in reviews~\cite{Herz08,Herz23}.
The appearance of these isomers is explained by occurrence in these nuclei of high-$K$ s-p
states near the Fermi level. Excitation energies of $8^-$ isomers
should be sensitive to the s-p spectra and pairing. Thus these isomers
are the robust test of the model. In our calculations,
$K^{\pi}=8^-$ isomers are QRPA states of electric multipolarity $\lambda\mu$=98,
 where $K=\mu$ and $\pi=(-1)^{\lambda}$.

The $8^-$ state at 1.254 MeV in $^{252}$No is usually assigned as neutron 2qp configuration
$nn[624\downarrow, 734\uparrow]$ \cite{Herz08,Herz23,Zhang_PRC25,Sul_EPJA07}. The same assignment is obtained
in our QRPA calculations for all four applied Skyrme forces, see Table~\ref{table4:8-}. Our calculations
predict this isomer as $F \to F+1$ excitation where $F$ is the Fermi level and $F+1$ is the next one.

\begin{table*} 
\caption{Characteristics of the lowest ($\nu$=1) calculated $8^-$ states in $^{252,254}$No: QRPA excitation
energies $E_{\nu}$, reduced transition probabilities $B(E98)$, main 2qp components $qq'$,
their energies $\epsilon_{qq'}$, contributions to the state norm $N_{qq'}$  and F-order.
For the convenience of the discussion, for $^{254}$No the second ($\nu$=2) $8^-$ states
are also depicted.}
\begin{tabular}{|c|c|c|c|c|c|c|}
\hline
Force & $E$ (MeV) & $B(E98)$ (W.u.)& $qq'$ & $\epsilon_{qq'}$ (MeV) & $N_{qq'}$ & F-order \\
\hline
 \multicolumn{7}{|c|}{$^{252}$No, $E_{\rm x}$=1.254 MeV} \\
\hline
SLy4 & 1.257 & 0.022   & $nn[624\downarrow,734\uparrow]$ & 1.211  & 0.998 & F,F+1 \\
SLy6 & 1.361 & 0.038  & $nn[624\downarrow,734\uparrow]$ & 1.317  & 0.996 & F,F+1 \\
SkM* & 1.330 & 0.025  & $nn[734\uparrow,624\downarrow]$ & 1.198  & 0.992 & F,F+1 \\
 SVbas & 1.913 & 0.119 &  $nn[624\downarrow,734\uparrow]$ & 1.751 & 0.912  & F,F+1\\
\hline
 \multicolumn{7}{|c|}{$^{254}$No, $E_{\rm x}$=1.295 MeV} \\
 \hline
SLy4 & 1.673 &  0.009  & $nn[734\uparrow,613\uparrow]$   & 1.695  & 0.996 & F,F+3 \\
     & 1.996 &  0.444  & $pp[514\downarrow,624\uparrow]$ & 1.994  & 0.847 & F+1, F+2 \\
  \cline{2-7}
SLy6 & 1.747 &  0.014  & $nn[734\uparrow,613\uparrow]$   & 1.780  & 0.994 & F,F+3 \\
     & 2.040 &  0.578  & $pp[514\downarrow,624\uparrow]$ & 1.966  & 0.993 & F+1, F+2 \\
   \cline{2-7}
SkM* & 1.554  & 0.333 & $pp[514\downarrow,624\uparrow]$  & 1.482 & 0.990 & F+1,F+2 \\
     & 1.645  & 0.0003 & $nn[734\uparrow,624\downarrow]$   & 1.610  & 0.985 & F-1,F \\
\cline{2-7}
 SVbas & 1.994 & 0.370 &  $pp[514\downarrow,624\uparrow]$ & 1.751 & 0.791  & F+1,F+2\\
      &  2.034 &   0.049  & $nn[734\uparrow,613\uparrow]$ & 2.026 & 0.811 & F,F+3 \\
\hline
\end{tabular}
\label{table4:8-}
\end{table*}

Following Table~\ref{table4:8-}, SLy4 very well describes the experimental $E_{\rm x}$=1.254
MeV of $8^-$ state in $^{252}$No. The forces SLy6 and SkM* slightly overestimate the
experimental values.  A much larger overestimation takes place for SVbas where the
 pairing is the most developed. In the case of the developed pairing,  the pairing blocking can decrease
 the calculated energies of 2qp states by 0.2-0.5 MeV~\cite{Jol11,Ne_PRC2016}. However,
 the calculations with the particle-number projection for rare-earth nuclei~\cite{Kuz_1986}
 show that the blocking effect too strongly suppresses the pairing in 2qp states.

Table~\ref{table4:8-} shows that QRPA $8^-$ state in $^{252}$No is basically formed by one 2qp component.
At the same time, the impact of residual interaction is also noticeable. Indeed, the collective shift
$\Delta E_{\rm coll}=E_{\nu=1}-\epsilon_{qq'}$ is 46, 44, 132 and 162 keV
for SLy4, SLy6, SkM* and SVbas, respectively.  The residual interaction upshifts the energy of
the lowest $8^-$ state as compared with minimal $\epsilon_{qq'}$. This means that IV
residual interaction overrides IS one.

The reduced transition probabilities $B(E98)$ for $8^-$ states with one dominant neutron 2qp pair are
obviously small since the electric transition operator (\ref{EL_operator}) includes only the proton part.
Instead, $8^-$ states with a large proton contribution are characterized by significant $B(E98)$
values. In $^{254}$No, the direct decay of $8^-$ isomer to $8^+$ member of the ground state band
was observed~\cite{Clark_PLB10,Wahid_PRC25}. However, this decay was not caused by an external electric
field and so cannot be considered as a justification of the proton structure of the $8^-$ isomer.
The deexitation of the neutron $8^-$ isomer can also occur through a measurable $\gamma$-decay.

Following Table~\ref{table4:8-}, the description of  $8^-$ state in $^{254}$No is much worse
than in $^{252}$No. The energy of this state is significantly overestimated by all four Skyrme forces.
Moreover, these forces predict three different 2qp configurations for $8^-$ state. The neutron assignment
from SLy4 and SLy6 looks more relevant since it treats $8^-$ excitation as 1ph transition $F \to F+3$.
Instead, SkM* and SVbas suggest the proton particle-particle transition $F+1 \to F+2$, which can occur
only at the condition of  well developed proton pairing. Note that our calculations for $^{254}$No
predict a suppression of the neutron pairing but not of the  proton one. Note also that
the neutron 1ph transition $624\downarrow \to 734\uparrow$, which gives $8^-$ state in
$^{252}$No, becomes particle-particle in  $^{254}$No and so is suppressed.

For the lowest $8^-$ state in $^{254}$No, we again see a noticeable impact of the residual
interaction. However now the collective shifts have different signs displaying a dominance of
IS interaction for SLy4 and SLy6 and of IV interaction for SkM* and SVbas.

The $8^-$ isomer in $^{254}$No was earlier investigated within various models, see e.g.
\cite{DN22,Sol91,Jol11,Ada_PRC2010,He_CPC2020}. The models~\cite{Sol91,Jol11} use Woods-Saxon single-particle
potential and schematic residual interaction, i.e. these models are not self-consistent. At the same time, they
implement the pairing blocking which allows to get a reasonable description of the energy of lowest 2qp $8^-$
state in $^{254}$No (1.4 MeV~\cite{Sol91} and 1.3 MeV~\cite{Jol11}). These models suggest different
assignments for the state: $nn[734\uparrow,613\uparrow]$ in Ref.~\cite{Sol91} and
$pp[514\downarrow,624\uparrow]$ in Ref.~\cite{Jol11}. These assignments correlate with our results in
Table~\ref{table4:8-}. The  Nilsson cranking model with the pairing projection~\cite{He_CPC2020}
predicts the assignment $pp[514\downarrow,624\uparrow]$ and suggests the state energy 1.272 MeV
in a good agreement with the experimental value. Altogether, studies~\cite{DN22,Sol91,Jol11,He_CPC2020}
show that the pairing blocking and projection are important for reproduction of $E_{\rm x}$. In particular,
this is confirmed by a good description of the energy of $8^-$ state by shell model with
projection-after-variation prescription\cite{DN22}.

Following previous \cite{Sol91,Jol11}  and our calculations, the different 2qp assignments of $8^-$ isomer
can be explained by close excitation energies of 2qp states $nn[734\uparrow,613\uparrow]$ and
$pp[514\downarrow,624\uparrow]$ in $^{254}$No.
So, even a modest difference  in the s-p spectra can result in different assignments.
Anyway, all the models predict that $8^-$ state in $^{254}$No is dominated by one 2qp configuration.
The calculations \cite{Sol91} show that, for this state, the coupling with complex configurations is negligible.
The two-center shell model ~\cite{Ada_PRC2010} gives the proton 2qp $8^-$ state at $\approx$1.27 MeV but here
a good description of the energy is obtained by a manual decrease of the pairing strength  by 15$\%$.

A more accurate 2qp assignment of the $8^-$ state in $^{254}$No could be done using experimentally
observed $\gamma$-decays to and from this state. However, even here we
have different conclusions. While the exploration~\cite{Hess_EPJA10} suggests the proton configuration
$pp[514\downarrow,624\uparrow]$, the study~\cite{Clark_PLB10} favors one of two neutron configurations,
 $nn[734\uparrow,613\uparrow]$ and $nn[734\uparrow,624\downarrow]$. The neutron assignments~\cite{Clark_PLB10}
 look more reasonable since they are argued  by the measured $\gamma$-decay from the neutron two-quasiparticle
 isomer $K^{\pi}=10^+$ $nn[734\uparrow,725\uparrow]$ located at 2.01 MeV to the
 rotational band built on the $K^{\pi}=8^-$ state.  Indeed, at the neutron assignments,
 the $\gamma$-decay involves the transition of a single neutron. Following our SLy6 calculations,
 this could be 0.5-MeV transition $725\uparrow (F+4) \to 613\uparrow (F+3)$
 or 3.4-MeV transition $725\uparrow (F+4) \to 624\uparrow (F-2)$. The former looks more realistic
 since its transition energy is closer to the experimental energy difference
 $E_{10^+}-E_{8^-}$=2.012-1.295=0.717 MeV. So,  the decay analysis~\cite{Clark_PLB10}
 confirms our SLy6 assignment $nn[734\uparrow,613\uparrow]$ shown in Table.~\ref{table4:8-}.
 Note also that, in a good agreement with results~\cite{Clark_PLB10}, our SLy6 calculations give
 the lowest 2qp neutron $K^{\pi}=10^+$ $nn[734\uparrow,725\uparrow]$
 state at 1.93 MeV, i.e. close to the experimental value 2.01 MeV.

The scenario~\cite{Clark_PLB10} was recently questioned by
measurements~\cite{Forge_JPCS2023,Forge_PhD}, where 2.01-MeV  $K^{\pi}=10^+$ band-head
was replaced by $K^{\pi}=11^-$ band-head   with the corresponding change of the parity
 of the members of the rotational band. Following our QRPA calculations, the lowest
 non-rotational state $K^{\pi}=11^-$ lies at  3.83 (SLy6), 4.04 (SkM*) and 3.51 (SVbas) MeV, i.e.
 much higher of the experimental energy 2.01 MeV. So, our calculations rather support the decay
 scheme~\cite{Clark_PLB10}.

\begin{table*}
\caption{Features of the lowest excited QRPA $0^+$ states in $^{252,254}$No: QRPA excitation
energies $E$, reduced transition probabilities $B(E20)$ and  $\rho^2(E0)$,
main  two 2qp components $qq'$, their energies $\epsilon_{qq'}$,
contributions to the state norm $N_{qq'}$  and F-order. For the sake of discussion, in $^{254}$No,
two lowest $0^+$ states are exhibited.}
\begin{tabular}{|c|c|c|c|c|c|c||c|}
\hline
Force &  $E$ (MeV) & $B(E20)$ (W.u.) & $\rho^2(E0)$ ($10^{-3}$)
& qq' & $\epsilon_{qq'}$ (MeV)& $N_{qq'}$ & F-order \\
\hline
& \multicolumn{7}{c|}{$^{252}$No} \\
\hline
SLy4    & 0.740 & 0.18 & 0.08  & $nn[734\uparrow, 734\uparrow]$     & 1.074 & 0.53 & F+1, F+1 \\
        &      &      &   & $nn[624\downarrow, 624\downarrow]$ & 1.348 & 0.30 & F,F \\
\hline
SLy6    & 0.774 & 0.02 & 0.29 & $nn[734\uparrow, 734\uparrow]$     & 1.070 & 0.58 & F+1, F+1 \\
        &      &      &   & $nn[624\downarrow, 624\downarrow]$ & 1.563 & 0.17 & F,F \\
\hline
SkM* &  0.838 & 1.12  & 1.24 & $pp[521\downarrow, 521\downarrow]$ &  1.014 & 0.46 & F,F \\
         &         &   &  & $pp[514\downarrow, 514\downarrow]$ & 1.093  & 0.42 & F+1,F+1  \\
\hline
SVbas & 1.249 & 6.41  & 2.50 & $pp[514\downarrow, 514\downarrow]$ &  1.215 & 0.56 & F+1,F+1 \\
           &      &   &  &  $pp[521\downarrow, 521\downarrow]$ & 1.186  & 0.36 & F,F\\
\hline
& \multicolumn{7}{c|}{$^{254}$No}\\
\hline
SLy4    & 0.616 & 0.08  &  0.08 & $nn[734\uparrow, 734\uparrow]$ & 1.021 & 0.51 & F, F \\
        &      &      &    & $nn[620\uparrow, 620\uparrow]$ & 1.156 & 0.25 & F+1,F+1 \\
        & 1.000 & 2.69   &  2.69 & $pp[514\downarrow, 514\downarrow]$ & 1.092 &0.57 &  F+1,F+1\\
        &       &       &     & $pp[521\downarrow, 521\downarrow]$ & 1.163 &0.30 &  F,F\\
\hline
 SLy6   & 0.224 & 0.002 & 0.01 & $nn[734\uparrow, 734\uparrow]$     & 1.048 & 0.41 & F, F \\
        &      &      &      & $nn[620\uparrow, 620\uparrow]$ & 1.267 & 0.27 & F+1,F+1 \\
        & 1.133 & 1.32 & 1.98 & $pp[514\downarrow, 514\downarrow]$ & 1.155 &0.56 &  F+1,F+1\\
        &       &       &     & $pp[521\downarrow, 521\downarrow]$ & 1.152 &0.45 &  F,F\\
 \hline
  SkM*  & 0.767 & 0.17 & 0.74 &  $nn[624\downarrow,624\downarrow]$ & 1.409  & 0.33 & F,F \\
        &       &      &      &  $nn[620\uparrow,620\uparrow]$ & 1.362 & 0.23  & F+1, F+1\\
        &  0.866 & 4.38 & 6.78 & $pp[521\downarrow, 521\downarrow]$ & 1.02 &0.45 &  F,F\\
           &  &  &  & $pp[514\downarrow, 514\downarrow]$ & 1.08 &0.43 &  F+1,F+1\\
\hline
   SVbas   & 1.236 & 6.36 & 2.54 &  $pp[514\downarrow,514\downarrow]$   & 1.083  & 0.43 & F+1,F+1 \\
           &     &   &  &  $pp[521\downarrow,521\downarrow]$         &  1.017 & 0.38  & F,F\\
           & 1.454 & 0.53  & 1.35 & $pp[633\uparrow, 633\uparrow]$ &  1.593 & 0.45 & F-1,F-1 \\
           &      &   &  &  $pp[521\downarrow, 521\downarrow]$       & 1.186  & 0.25 & F,F\\
\hline
\end{tabular}
\label{table5:QRPA_0+}
\end{table*}

 The issue has become even more intriguing after publication of a new high-resolution
 $\gamma$-decay data for $^{254}$No~\cite{Wahid_PRC25}, which confirm and additionally specify
 the previous decay scheme~\cite{Clark_PLB10}. However, unlike the previous study~\cite{Clark_PLB10},
  the proton assignment $pp[514\downarrow,624\uparrow]$ was suggested for $8^-$ isomer.
 This assignment was based on the calculation results ("universal" WS potential + LN procedure for the pairing)
 with the lowest proton 2qp configuration. Besides, this configuration was justified by decay
 of $8^-$ isomer to the rotational band of proton isomer $K^{\pi}=3^+$. Actually, both neutron~\cite{Clark_PLB10}
 and proton~\cite{Wahid_PRC25} assignments were justified by similar arguments but using
 different parts of the decay chain, the decay $10^+ \to 8^-$ in the analysis~\cite{Clark_PLB10}
 and decay  $8^- \to 3^+$ in the discussion~\cite{Wahid_PRC25}.  We incline
 to arguments~\cite{Clark_PLB10} since  then we deal with $\Delta K=$2 transition
 which should be much stronger than  $\Delta K=$5 one.
 Our SLy4 and SLy6 results for $8^-$-isomer just confirm the neutron assignment  $nn[734\uparrow,613\uparrow]$.

 As seen from Table~\ref{table4:8-},  the energy interval between the lowest neutron and proton
 2qp $8^-$ configurations in $^{254}$No can be rather small, e.g.  186 keV for SLy6
 and 128 keV for SkM*. These values are comparable with collective shifts mentioned above. So, in principle,
one cannot exclude the mixture of the neutron and proton $8^-$ 2qp pairs by the residual $\lambda\mu$=98
 interaction. This could partly reconcile two optional assignments~\cite{Clark_PLB10}
 and~\cite{Wahid_PRC25} for $8^-$ isomer in $^{254}$No.

 To our opinion, just well established isomers $8^-$ and $2^-$ in  $^{252}$No and $3^+$ in
 $^{254}$No have to be used for the primary testing of the theory. Moreover, only a simultaneous
 description of these isomers can justify the theory validity. To our knowledge,
 there was no yet theoretical studies which successfully describe this set of isomer's data.
 As for the disputed $8^-$ isomer in $^{254}$No,  its description demands very fine tuning of the
 s-p scheme and pairing and so it is worth to use this isomer only for the secondary fine tuning.
 In Appendix A we show that  alternative Skyrme parametrizations (UNEDF1~\cite{Kor_UNEDF1},
 UNEDF1$^{\rm SO}$~\cite{Shi_PRC2014}  and UNEDF2~\cite{Kor_UNEDF2}) do not provide the simultaneous
 description of the isomers and, in this sense, demonstrate a worse performance than SLy6.

The above analysis as well as a good SLy6 description of $3^+$ and $8^-$ in $^{252}$No and $2^-$
state in $^{254}$No (see below) show that this force is reasonable enough for the thorough inspection
 of the low-energy spectra in $^{252,254}$No. The force SLy4 supplements SLy6 by correcting
 its results in the case of a weak neutron pairing in $^{254}$No.

\subsection{Pairing vibrations $K^{\pi}=0^+$}

As mentioned above, our calculations predict a reduction of the neutron pairing in $^{254}$No.
This should lead to a significant downshift of the energy of $K^{\pi}=0^+$ pairing
vibrational states with a dominant  neutron structure.  Table~\ref{table5:QRPA_0+}
shows that this is indeed the case for SLy4, SLy6 and  SkM* where the neutron pairing is suppressed.
These forces give first $0^+_1$ states below 1 MeV. The lowest energy 0.224 MeV is obtained
for SLy6 where  we have a collapse of the neutron pairing. Obviously, so low energy is an
 artifact of the BCS description of a weak pairing. In this connection, SLy4 energy 0.611 MeV
 for $0^+_1$ state is more realistic (here we have a drop but not collapse of the neutron
 pairing). In $^{252}$No, the forces SLy4, SLy6 and SkM* also give $0^+_1$-energies
 below 1 MeV. In both $^{252}$No and $^{254}$No, the $0^+_1$  states are collective (the largest
2qp component exhausts only 33-58$\%$ of the state norm) and dominated by
neutron contributions. The dominant neutron structure leads to small values of $B(E2)$ and
$\rho^2(E0)$. In $^{254}$No, the forces SLy4, SLy6 and especially SkM*
also provide  second  $0^+_2$ low-energy states.

For the forces with more developed neutron pairing,  we get $0^+_1$ states with basically proton
structure (SkM* for $^{252}$No and SVbas for $^{252,254}$No). The states are also collective but,
because of the proton structure, exhibit large $B(E2)$ and $\rho^2(E0)$ values.

The $\rho^2(E0)$ values for $^{252,254}$No in
Table~\ref{table5:QRPA_0+} are smaller than those in $^{282}$Cn where they reach
$\approx 21 \cdot 10^{-3}$ \cite{Samark_PRC23}. At the same time, our $\rho^2(E0)$ values
correspond to QPM estimations for rare-earth and actinide nuclei~\cite{GS_book74,Sol_book76,ISS05}.

The $0^+$ states in Table~\ref{table5:QRPA_0+} are mainly composed of {\it diagonal} 2qp configurations,
which means that they are  {\it pairing vibrations}. Such  states are well known in even-even deformed nuclei in rare-earth~\cite{GS_book74,Sol_book76,SSS96,ISS05,IS08} and actinide \cite{ISS05} regions.
The most thorough analysis for low-energy $0^+$ states was performed within QPM~\cite{Sol_book76}.
The lowest excited $0^+$ states usually have the energy around 1 MeV.
Their reduced transition probabilities $B(E20, 0^+0_{\rm gs} \to 2^+0)$  depend on the
pairing strength and contribution of $\beta$-vibrations.
Pairing vibrational states can be observed in $(p,t)$ and $(t,p)$ transfer reactions~\cite{GS_book74,Sol_book76}.
Following QPM studies, the lowest  $K^{\pi}_{\nu}=0^+_1$ state usually has negligible two-phonon
admixtures and so can be reasonably described within QRPA. One should distinguish
pairing vibrational states  from the low-lying $0^+$ states arising from a shape coexistence~\cite{De06,ER00}.
The description of the latters is beyond  the scope of QPM and present QRPA method.

In the recent experiment for $^{254}$No~\cite{Forge_JPCS2023}, an excited
$0^+$ state at 0.888 MeV was identified. For the moment, this is the lowest observed non-rotational state
in the multipole spectrum of $^{254}$No. Our SLy4 and SkM* results in Table~\ref{table5:QRPA_0+}
reasonably reproduce the energy of this state (for SkM* both first and second $0^+$ states can pretend to
describe the experiment). Following the theoretical predictions~\cite{De06,ER00}, the authors of
Ref.~\cite{Forge_JPCS2023} treat the observed $0^+$ state as a consequence of the coexistence of
normally deformed and superdeformed  shapes. The impact of the prolate and triaxial shapes is also
admitted.

Our SLy4, SLy6, SkM*  and SVbas calculations also predict for $^{254}$No two deep minima in PES:
the main minimum at "normal" deformation $\beta \approx$ 0.3 and the second
minimum at superdeformation $\beta \approx$ 0.9-1.0. However,
our SkM* and SVbas calculations performed with $\beta \approx$ 0.3 rather well reproduce the experimental
energies of low-spin states of the g.s. rotational band  in $^{254}$No or somewhat
underestimate them (SLy4, SLy6), see e.g. Figs.~\ref{fig5:252No_QRPA} and \ref{fig6:254No_QRPA} below.
Instead, if we use the superdeformation $\beta \approx$ 0.9-1.0,
the moments of inertia become much larger and  agreement with the experiment
rotational spectrum drastically worsens. Note also that rotational bands for the superdeformed state
were not yet observed in $^{254}$No. Moreover, the existence of the superdeformed state as such is
questionable as it is prone to be unstable against fission.
Altogether, to our opinion, the coupling of "normal" and superdeformed
shapes in $^{254}$No is too weak to affect noticeably the low-energy spectrum.
So we can safely consider only the prolate g.s. at $\beta\approx$0.3. Following these arguments,
our explanation of 0.888-MeV state as a pairing vibration seems to be the most plausible.

Recent shell-model calculations with projection after variation~\cite{DN22} give
in $^{254}$No the  $K^{\pi}=0^+$ state at 0.86 MeV. However, the authors do not analyze
the origin and features of the state.

Low-energy $0^+$ states were earlier predicted within triaxial beyond-mean-field (TBMF)
prescription in Fl ($Z$=114)~\cite{Egido_PRL20} and Cn ($Z$=112)~\cite{Samark_PRL21,Samark_PRC23}
isotopes. Their excitation energies were suggested at 0.4-1.3 MeV in
$^{288-292}$Fl~\cite{Egido_PRL20} and 0.7-2.2 MeV in  $^{282-288}$Cn~\cite{Samark_PRL21}.
Moreover, the excited  $0^+$ state at $E_{\rm x}\approx$0.6 MeV was experimentally observed in
$^{282}$Cn~\cite{Samark_PRL21,Samark_PRC23} within the coincidence experiment (population of this state through
$\alpha$-decay from $^{286}$Fl and subsequent deexcitation of the state through electron
internal conversion). Perhaps, all these excited $0^+$ states
are also basically pairing vibrational.

Unlike Fl and Cn isotopes with Z equal or close to the suggested spherical magic number
$Z$=114, our calculations concern the case of the suppressed pairing in the mid of
the chain of {\it well deformed} nobelium isotopes whose $Z$ and $N$ are far from spherical
magic numbers. To the best of our knowledge,  previous studies of $0^+$ states in well deformed nuclei
did not consider the cases of the suppressed pairing.

The excited $0^+$ state can serve as a sensitive indicator of the pairing
in $^{254}$No. It would be interesting to look for this state in the coincidence experiments
like those in Refs.~\cite{Samark_PRL21,Samark_PRC23}. Besides, the exploration of
this state in photonuclear and $(p,t)$ reactions could give an important information whether
we indeed have a significant drop of the neutron pairing in $^{254}$No. If so, then this would
confirm the existence of the shell gap in the neutron s-p spectrum of this nucleus.

\begin{table} 
\caption{Features of two lowest ($\nu$=1,2) QRPA $K^{\pi}=2^+$ states in $^{252,254}$No:
excitation energies $E$ (in MeV), reduced transition probabilities
$B(E22)$ (in W.u.), main 2qp components $qq'$, their energies $\epsilon_{qq'}$ (in MeV),
contributions to the state norm $N_{qq'}$  and F-order.}
\begin{tabular}{|c|c|c|c|c|c|c|c|}
\hline
Force &  $E$ & $B(E22)$ & qq' & $\epsilon_{qq'}$ & $N_{qq'}$ & F-order \\
\hline
& \multicolumn{6}{|c|}{$^{252}$No} \\
\hline
SLy6 & 1.58 &  3.87 & $nn[622\downarrow, 620\uparrow]$ & 2.33  & 0.39 & F+2,F+2 \\
     &      &       & $pp[521\downarrow, 521\uparrow]$ & 2.06  & 0.32 & F,F-2 \\
\cline{2-7}
    & 2.08   &   0.001   & $pp[514\downarrow, 521\uparrow]$ & 2.06  & 1.00 & F+1,F-2\\
\hline
%
SkM* & 1.70   &  0.06 &  $pp[512\uparrow, 521\downarrow]$ & 1.61  &  0.99 & F+3,F\\
\cline{2-7}
    & 1.78   &   2.71   & $nn[622\uparrow, 620\uparrow]$ & 2.28  &  0.35 & F-1,F+2\\
    &        &      & $nn[624\downarrow, 622\downarrow]$ & 2.14  & 0.29 & F+1,F+3\\
\hline
%
SVbas & 1.62  &  2.72   & $pp[521\uparrow, 521\downarrow]$ & 1.95  & 0.38 & F-2,F\\
      &       &      & $nn[622\uparrow, 620\uparrow]$     &  2.48 & 0.29 & F-1,F+2\\
\cline{2-7}
       & 1.89   &  0.004 &  $pp[512\uparrow, 521\downarrow]$ & 1.86  &  0.99 & F+3,F\\
\hline
& \multicolumn{6}{|c|}{$^{254}$No}\\
\hline
SLy6     & 1.31 &  0.17 & $nn[622\uparrow,620\uparrow]$ &  1.32 & 0.97 & F-1,F+1\\
\cline{2-7}
         & 1.53 &  3.71    & $nn[622\uparrow,620\uparrow]$ &  2.24 & 0.42 & F-1,F+1\\
         &      &       & $pp[521\uparrow, 521\downarrow]$ & 2.05  & 0.27 & F-2,F\\
\hline
SkM* &  1.32 & 2.62  &  $nn[624\downarrow, 622\downarrow]$ & 1.63  &  0.60 & F,F+2\\
    &        &      & $nn[622\downarrow, 620\uparrow]$ & 1.60  &  0.18 & F+2,F+1\\
\cline{2-7}
    &   1.62 &   0.015   & $nn[622\downarrow, 620\uparrow]$ & 1.60  & 0.80 & F+2,F+1\\
\hline
SVbas &  1.45 & 4.46 &  $nn[622\downarrow,620\uparrow]$ & 1.77  &  0.40 & F+2,F+1\\
        &      &       & $pp[521\uparrow, 521\downarrow]$ & 1.95  & 0.20 & F-2,F\\
 \cline{2-7}
    &   1.87 &   0.32   & $nn[622\downarrow, 620\uparrow]$ & 1.77  & 0.56 & F+1,F+2\\
     &      &       & $pp[521\uparrow, 521\downarrow]$ & 1.95  & 0.21 & F-2,F\\
\hline
\end{tabular}
\label{table6:2+}
\end{table}

\begin{table} 
\caption{The same as in Table~\ref{table6:2+} but for $K^{\pi}=3^+$. For $^{254}$No,
the SLy4 results are also shown.}
\begin{tabular}{|c|c|c|c|c|c|c|c|}
\hline
Force &  $E$ & $B(E43)$ & qq' & $\epsilon_{qq'}$ & $N_{qq'}$ & F-order \\
\hline
& \multicolumn{6}{|c|}{$^{252}$No} \\
\hline
SLy6 & 1.10 &  3.04  & $pp[521\downarrow, 514\downarrow]$ & 1.16 & 0.99 & F,F+1 \\
\cline{2-7}
    & 2.13   &   2.91   & $pp[521\downarrow, 512\uparrow]$ & 2.11  & 0.95 & F,F+3\\
\hline
%
SkM* & 1.00   &  3.61 &  $pp[521\downarrow, 514\downarrow]$ & 1.05  &  0.97 & F,F+1\\
\cline{2-7}
    & 1.69   &   2.67   & $pp[521\downarrow, 512\uparrow]$ & 1.61  &  0.94 & F,F+3\\
\hline
%
SVbas & 1.19  &  2.73   & $pp[521\downarrow, 514\downarrow]$ & 1.21  & 0.98 & F,F+1\\
   \cline{2-7}
       & 1.93   &  2.43 &  $pp[521\downarrow, 512\uparrow]$ & 1.86  &  0.95 & F,F+3\\
\hline
& \multicolumn{6}{|c|}{$^{254}$No, $E_{\rm x}$=0.987 MeV}\\
\hline
SLy4     & 1.07 &  3.21 & $pp[521\downarrow, 514\downarrow]$ &  1.13 & 0.99 & F,F+1\\
\cline{2-7}
         & 1.74 &  0.67     & $nn[620\uparrow, 613\uparrow]$ &  1.76 & 0.96 & F+1,F+3\\
\hline
SLy6     & 1.11 &  2.41 & $pp[521\downarrow, 514\downarrow]$ &  1.15 & 0.99 & F,F+1\\
\cline{2-7}
         & 1.89 &  1.78     & $nn[620\uparrow, 613\uparrow]$ &  1.89 & 1.00 & F+1,F+3\\
\hline
SkM* &  1.01 & 3.24 &  $pp[521\downarrow, 514\downarrow]$ & 1.05  &  0.97 & F,F+1\\
\cline{2-7}
    &   1.41 &   2.15  & $nn[624\downarrow, 620\uparrow]$ & 1.39  & 1.00 & F,F+1\\
\hline
SVbas &  1.17 & 3.00  &  $pp[521\downarrow, 514\downarrow]$ & 1.20  &  0.99 & F,F+1\\
    \cline{2-7}
    &   1.87 &   3.28   & $nn[620\uparrow, 613\uparrow]$ & 1.98  & 0.48 & F+1,F+3\\
     &      &       & $pp[521\downarrow, 512\uparrow]$ & 1.89  & 0.47 & F,F+3\\
\hline
\end{tabular}
\label{table7:3+}
\end{table}

\begin{table} 
\caption{The same as in Table~\ref{table6:2+} but for $K^{\pi}=4^+$.}
\begin{tabular}{|c|c|c|c|c|c|c|c|}
\hline
Force &  $E$ & $B(E44)$ & qq' & $\epsilon_{qq'}$ & $N_{qq'}$ & F-order \\
\hline
& \multicolumn{6}{|c|}{$^{252}$No} \\
\hline
SLy6 & 1.16 &  6 $10^{-4}$  & $pp[521\downarrow, 514\downarrow]$ & 1.16 & 1.00 & F,F+1 \\
\cline{2-7}
    & 2.11   &   1.78   & $nn[624\downarrow, 620\uparrow]$ & 2.34  & 0.50 & F,F+2\\
    &        &        & $nn[622\uparrow, 622\downarrow]$ & 2.41  & 0.42 & F-1,F+3\\
\hline
%
SkM* & 1.08   &  0.04 &  $pp[521\downarrow, 514\downarrow]$ & 1.05  &  1.00 & F,F+1\\
\cline{2-7}
    & 1.85   &  0.37  & $nn[624\downarrow, 620\uparrow]$ & 1.87  &  0.98 & F+1,F+2\\
\hline
%
SVbas & 1.22  &  0.04   & $pp[521\downarrow, 514\downarrow]$ & 1.21  & 1.00 & F,F+1\\
   \cline{2-7}
       & 2.06   &  0.86 & $nn[624\downarrow, 620\uparrow]$  & 2.09  &  0.93 & F,F+2\\
\hline
& \multicolumn{6}{|c|}{$^{254}$No, $E_{\rm x}$=1.203 MeV}\\
\hline
SLy4     & 1.14 &  0.05 & $pp[521\downarrow, 514\downarrow]$ &  1.13 & 1.00 & F,F+1\\
\cline{2-7}
         & 1.77 &   0.002  & $nn[620\uparrow, 613\uparrow]$ &  1.76 & 1.00 & F+1,F+3\\
\hline
SLy6     & 1.16 &  0.07 & $pp[521\downarrow, 514\downarrow]$ &  1.15 & 1.00 & F,F+1\\
\cline{2-7}
         & 1.89 &   $10^{-4}$   & $nn[620\uparrow, 613\uparrow]$ &  1.89 & 1.00 & F+1,F+3\\
\hline
SkM* &  1.07 & 0.05 &  $pp[521\downarrow, 514\downarrow]$ & 1.05  &  1.00 & F,F+1\\
\cline{2-7}
    &   1.36 &   0.47  & $nn[624\downarrow, 620\uparrow]$ & 1.39  & 0.99 & F,F+1\\
\hline
SVbas &  1.21 & 0.04  &  $pp[521\downarrow, 514\downarrow]$ & 1.20  &  1.00 & F,F+1\\
    \cline{2-7}
    &   1.98 &   0.06   & $nn[620\uparrow, 613\uparrow]$ & 1.98  & 0.94 & F+1,F+3\\
\hline
\end{tabular}
\label{table8:4+}
\end{table}

\begin{table*}
\caption{Features of the lowest QRPA octupole states with $K^{\pi}=0^-, 1^-, 2^-, 3^-$
in $^{252,254}$No: excitation energies $E$ (in MeV), reduced transition probabilities
$B(E3K)$ (in W.u.), main 2qp components $qq'$, their energies $\epsilon_{qq'}$ (in MeV),
contributions to the state norm $N_{qq'}$  and F-order.}
\begin{tabular}{|c||c|c|c|c|c|c|c||c|c|c|c|c|c|c|}
\hline
Force & K$^\pi$ & $E$ & $B(E3K)$ &  qq' & $\epsilon_{qq'}$ & $N_{qq'}$ & F-order
 & K$^\pi$ & $E$ & $B(E3K)$ &  qq' & $\epsilon_{qq'}$ & $N_{qq'}$ & F-order\\
\hline
& \multicolumn{7}{c||}{$^{252}$No} & \multicolumn{7}{c|}{$^{254}$No}\\
\hline
SLy6 & 0$^-$ &    1.24   &  9.1 & $pp[514\downarrow, 633\uparrow]$ & 1.35  & 0.93 & F+1,F-1
    & 0$^-$ & 1.25 &  11.2  & $pp[514\downarrow,633\uparrow]$ & 1.38  & 0.87 &F+1,F-1\\
\cline{2-15}
& 1$^-$ & 1.41 &  1.5 & $nn[734\uparrow, 624\downarrow]$ & 1.32  &  0.98 & F+1,F
&1$^-$ &  1.54 &  8.4 & $nn[734\uparrow,613\uparrow]$ & 1.78  & 0.82 & F,F+3\\
\cline{2-15}
& 2$^-$ & 0.95 &  11.5 & $nn[622\uparrow, 734\uparrow]$ &  1.30 & 0.92 & F-1,F+1
& 2$^-$ & 2.12 &  0.6 & $nn[622\uparrow, 734\uparrow]$ & 2.13  & 0.94 & F-1,F\\
\cline{2-15}
& 3$^-$ &   1.35    &  0.1 & $pp[633\uparrow, 521\downarrow]$ & 1.35  & 1.00 & F-1,F
& 3$^-$ & 1.28 &  0.03 & $nn[734\uparrow,622\downarrow]$ & 1.213  & 0.94 & F,F+2\\
\hline
\hline
SkM* & 0$^-$ &    1.35   &  20.7 & $pp[514\downarrow, 633\uparrow]$ & 1.52  & 0.79 &F+1, F-1
& 0$^-$ & 1.37 &  16.3 & $pp[514\downarrow,633\uparrow]$ & 1.51  & 0.84 & F+1,F-1\\
\cline{2-15}
& 1$^-$ & 1.16 &  2.2 & $nn[734\uparrow, 624\downarrow]$ & 1.20  &  0.97 & F,F+1
& 1$^-$ &  1.47 &  1.5 & $pp[624\uparrow,514\downarrow]$ & 1.48  & 0.95 & F+2,F+1\\
\cline{2-15}
& 2$^-$ & 1.46 &  6.2 & $nn[734\uparrow, 622\uparrow]$ & 1.61 & 0.92 & F,F-1
& 2$^-$ & 1.80 &  3.7 & $nn[725\uparrow,624\downarrow]$ & 1.71  & 0.85 & F+3,F\\
\cline{2-15}
& 3$^-$ & 1.48 &  0.05 & $pp[633\uparrow, 521\downarrow]$ & 1.48 & 1.00& F-1,F
& 3$^-$ & 1.48 &  0.04 & $pp[633\uparrow,521\downarrow]$ & 1.48 & 1.00 & F-1,F\\
\hline
\hline
SVbas &0$^-$ &  1.32   &  7.7 & $pp[514\downarrow, 633\uparrow]$ & 1.42  & 0.92 & F+1, F-1
& 0$^-$ & 1.30 &  7.4 & $pp[514\downarrow,633\uparrow]$ & 1.40  & 0.92 & F+1,F-1\\
\cline{2-15}
&1$^-$ & 1.71 &  6.1 & $nn[734\uparrow, 624\downarrow]$ & 1.75  &  0.77 & F+1,F
& 1$^-$ &  1.72 &  12.3 & $nn[734\uparrow,613\uparrow]$ & 2.03  & 0.42 & F,F+3\\
&  &  &   & $pp[633\uparrow, 512\uparrow]$ & 2.06  &  0.10 & F-1,F+3
&   &   &   & $pp[633\uparrow,512\uparrow]$ & 2.09  & 0.30 & F-1,F+3\\
\cline{2-15}
&2$^-$ & 1.62 &  12.6 & $nn[734\uparrow,622\uparrow]$ & 1.9 & 0.72 & F+1,F-1
& 2$^-$ & 1.90 &  14.5 & $pp[633\uparrow,521\uparrow]$ & 2.15 & 0.44 & F-1,F\\
&  &  &   & $pp[633\uparrow,521\uparrow]$ & 2.15 & 0.13 & F-1,F
& & &    & $nn[734\uparrow,622\uparrow]$ & 2.33  &  0.26 & F,F-2\\
\cline{2-15}
& 3$^-$ &   1.40  &  0.06 & $pp[633\uparrow,521\downarrow]$ & 1.40  & 1.00 &F-1,F
 &3$^-$ & 1.39 & 0.05 & $pp[633\uparrow,521\downarrow]$ & 1.40  & 1.00 & F-1,F\\
\hline
\end{tabular}
\label{table9:octupoles}
\end{table*}

\subsection{Quadrupole states  $K^{\pi}=2^+$}

As seen in Table~\ref{table6:2+}, SLy6, SkM* and SVbas give similar predictions for the
 energies of the first (1.58-1.70 MeV in $^{252}$No and 1.31-1.45 MeV in $^{254}$No)
 and second (1.78-2.08 MeV in $^{252}$No and 1.53-1.87 MeV in $^{254}$No)
$K^{\pi}=2^+$  states. At the same time, these forces suggest
 essentially different structure of the states. The first $K^{\pi}=2^+$  states
 are $\gamma$-vibrational collective for SLy6 and SVbas in $^{252}$No and for SkM* and SVbas in $^{254}$No.
 Instead, the first $2^+$ states  are purely 2qp for SkM* in $^{252}$No and SLy6 in $^{254}$No.
 In most of the cases, if the first state is collective, then the next one is 2qp
 and vice versa. All the calculated $2^+$ states lie above
the observed $2^-$ ($^{252}$No) and $3^+$ ($^{254}$No) $K$-isomers.

In contrast to our results, IBM calculations~\cite{Ef-Iz_PAN2021}
predict first $K^{\pi}=2^+$ states at much lower energies 1.09 MeV ($^{252}$No) and 0.94 MeV ($^{254}$No).
The relevance of the various theoretical predictions for No isotopes can be discriminated by
future  experiments.

\subsection{Hexadecapole states with $K^{\pi}=3^+$ and $4^+$}

In Table~~\ref{table7:3+}, the calculated features of two lowest hexadecapole $K^{\pi}=3^+$
states in $^{252,254}$No are collected. It is seen that all the applied Skyrme forces
rather well describe the excitation energy $E_{\rm x}$=0.987 MeV of the observed $3^+$ state in
$^{254}$No~\cite{Herz08,Herz_Nature2006,Tan_PRL2006,Clark_PLB10}.
The first $3^+$ state is purely 2qp and, in agreement with previous studies~\cite{Herz08,DN24,Herz_Nature2006,Tan_PRL2006,Clark_PLB10,Jol11,He_CPC2020,Liu11},
is assigned as $pp[521\downarrow, 514\downarrow]$. This confirms the relevance of
our single-particle schemes.

 For $^{252}$No, our calculation also give purely 2qp $3^+$ state
 at a similar energy 1.00-1.19 MeV and with the same assignment
 $pp[521\downarrow, 514\downarrow]$.

In Table~~\ref{table7:3+}, the second $3^+$ state is also 2qp (with exception of SVbas case in
$^{254}$No).  So, the impact of the hexadecapole residual interaction on $3^+$ states
in $^{252,254}$No looks negligible. Instead, QPM calculations~\cite{Jol11} predict a noticeable impact
of the hexadecapole residual interaction in nobelium isotopes. Besides, a significant impact
of the residual interaction was found in QPM calculations for hexadecapole states in
rare-earth nuclei~\cite{IS08,Ne86}.

Table~\ref{table8:4+} shows that the calculated first $K^{\pi}=4^+$ states in $^{252,254}$No have the
energies and structure very similar to $3^+$ states. This is not surprising since
both $3^+$ and $4^+$ states are basically formed by the same proton 2qp configuration
$pp[521\downarrow, 514\downarrow]$  with $|K_1-K_2|$=3 and $K_1+K_2$=4, respectively. This means that,
near the observed $3^+$ state in $^{254}$No, there should be some $4^+$ state. Indeed, in the recent
measurements~\cite{Forge_JPCS2023}, a $4^+$ state was observed in $^{254}$No at 1.203 MeV, i.e. rather
close to our predictions.
In the shell-model calculations~\cite{DN22}, the $4^+$ state in in $^{254}$No is predicted at 1.250 MeV.

Following our calculations, a doublet $3^+, 4^+$ is expected in $^{252}$No as well.
In both $^{252}$No  and $^{254}$No, the band heads  $3^+$ and $4^+$ and
corresponding rotational bands should exhibit a strong Coriolis coupling.

As shown in Table~\ref{table8:4+}, the pair $pp[521\downarrow, 514\downarrow]$ has a
feature $B(E43) \gg B(E44)$. This is probably caused by violation in this configuration
of the Nilsson selection rule $\Delta K=\Delta \Lambda$~\cite{Sol_book76,MN59} for E44 transitions.

\begin{figure*}[h] 
\centering
\includegraphics[width=1.0\textwidth]{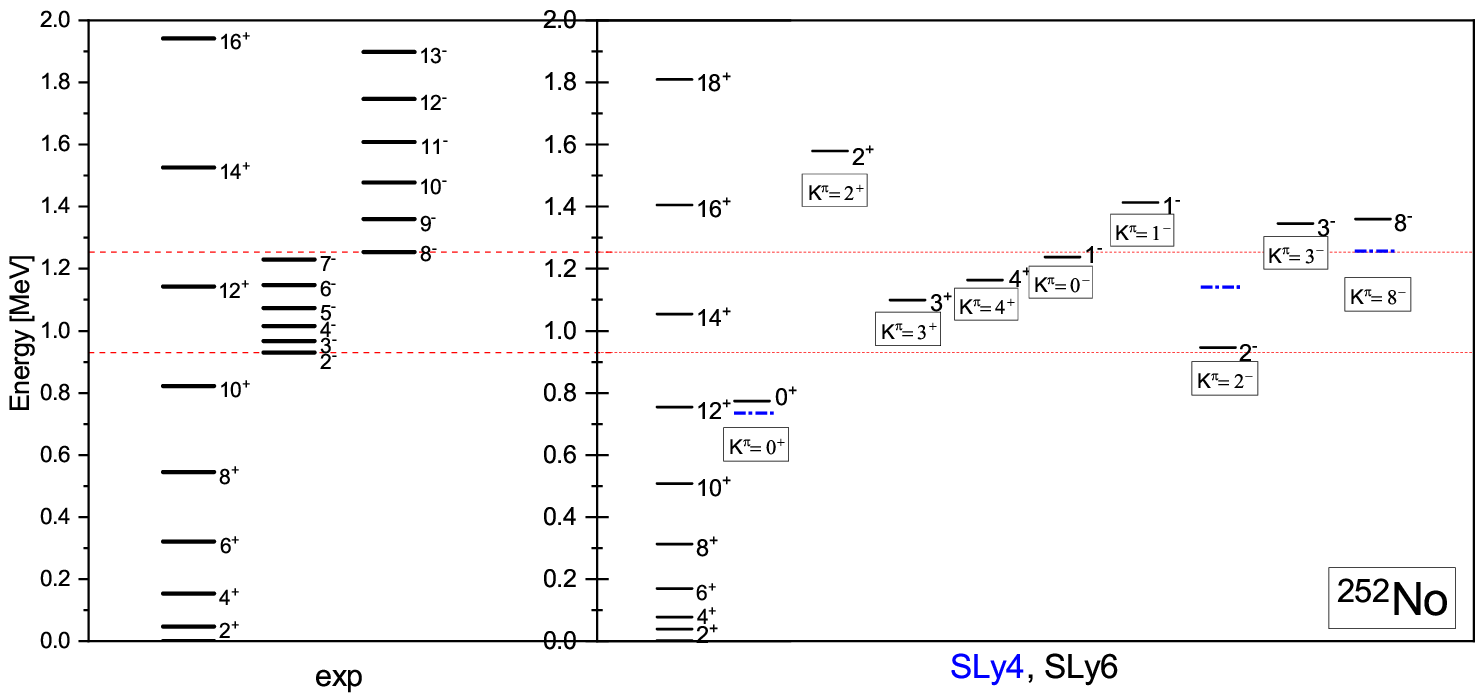}
\caption{Experimental~\cite{nndc}  (left) and QRPA SLy6 and SLy4 (right) spectra in $^{252}$No.
The state energies are shown by black solid (SLy6) and  blue dash-short (SLy4) lines.
The rotational states are marked  by $I^{\pi}$ on the right. In the right panel, the band
heads are marked by boxed $K^{\pi}$. For $K^{\pi}=0^-$, the lowest $I^{\pi}=1^-$
state is shown. The horizontal dashed red lines mark the experimental energies 0.929 MeV of
$K^{\pi}=2^-$ state and 1.254 MeV of $K^{\pi}=8^-$ state.}
\label{fig5:252No_QRPA}
\end{figure*}
\begin{figure*}[h] 
\centering
\includegraphics[width=1.0\textwidth]{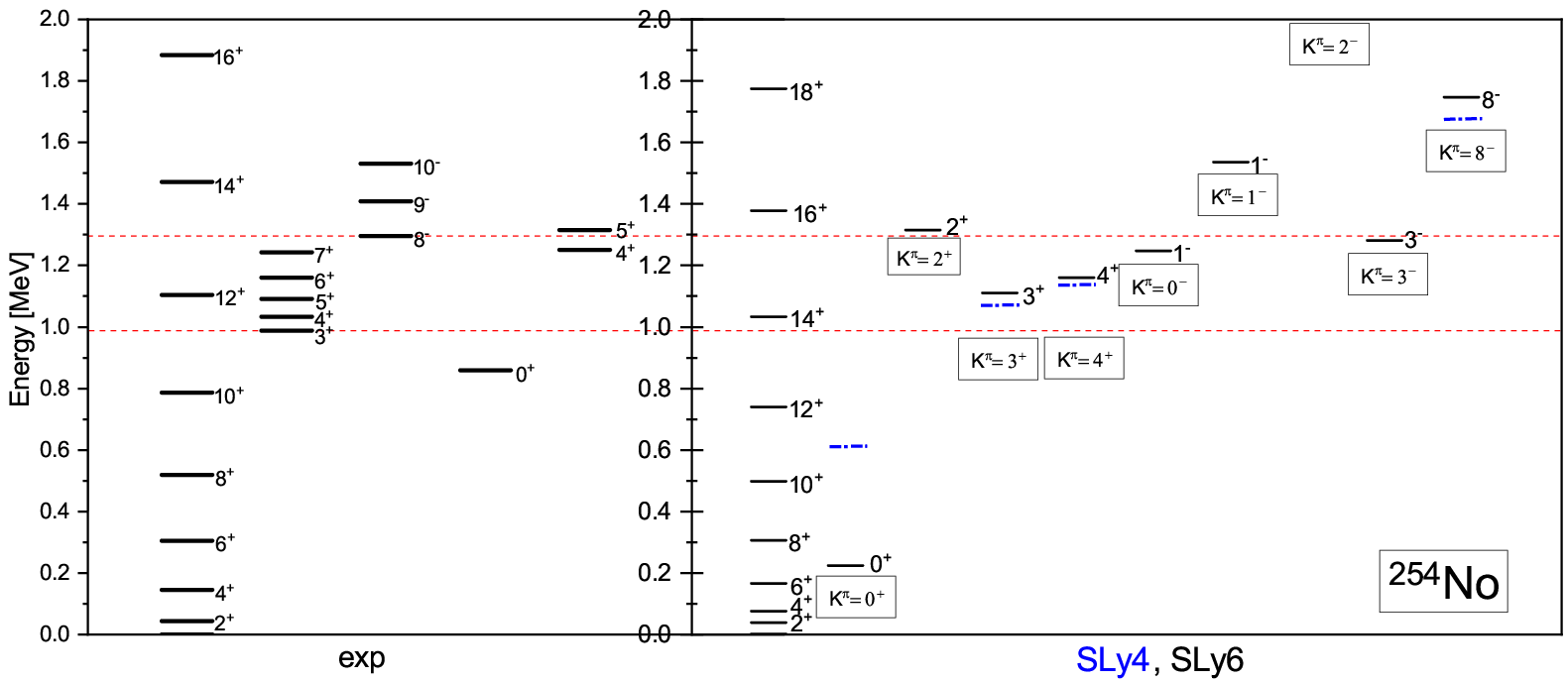}
\caption{The same as in fig.~\ref{fig5:252No_QRPA} but for $^{254}$No. The horizontal
dashed red lines mark the experimental energies 0.987 MeV of $K^{\pi}=3^+$ state and 1.295 MeV of
$K^{\pi}=8^-$ state. The energy E=2.12 MeV of the lowest SLy6 $K^{\pi}=2^-$ state is beyond
the frame of the figure.}
\label{fig6:254No_QRPA}
\end{figure*}
\begin{figure*}
\centering
\includegraphics[width=0.99\textwidth]{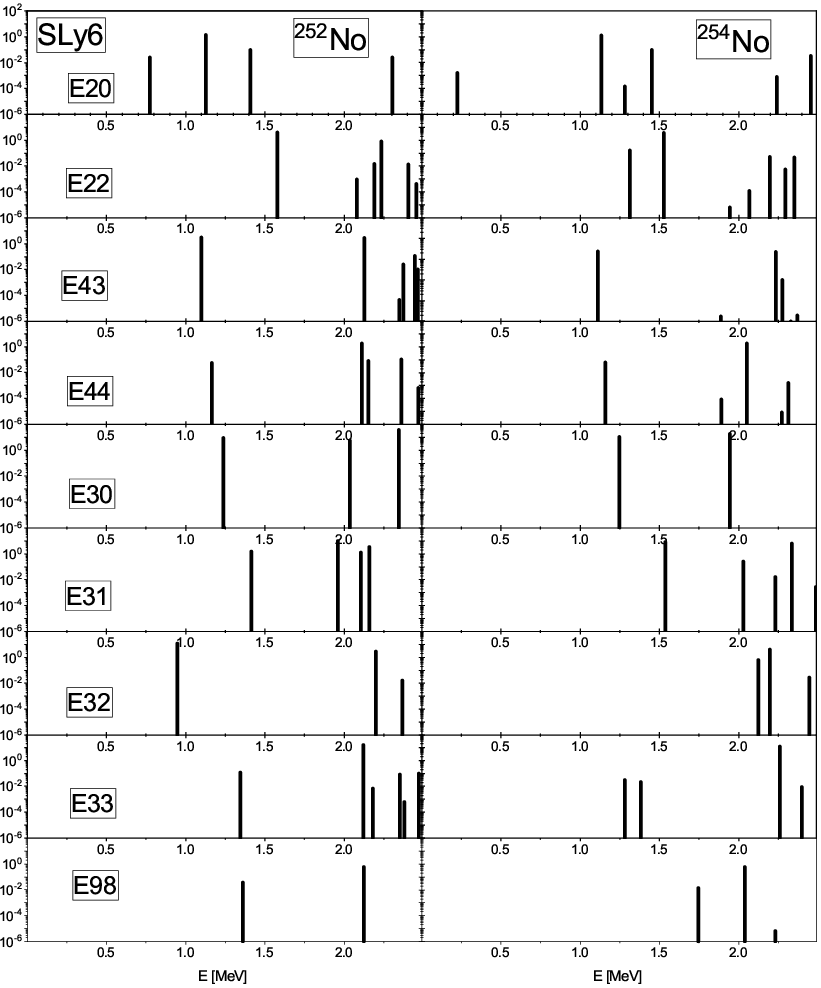}
\caption{SLy6 reduced transition probabilities $B(E\lambda\mu)$ (W.u. in log scale) for low-energy (E $<$ 2.5 MeV)
multipole QRPA states in $^{252}$No and $^{254}$No.}
\label{fig7:B(EL)}
\end{figure*}

\subsection{Octupole states with  $K^{\pi}=0^-, 1^-, 2^-$ and $3^-$}

Our results for octupole states in $^{252,254}$No are exhibited in
Table~\ref{table9:octupoles}. The force SLy6 gives for $2^-$ isomer in $^{252}$No
the energy $E$=0.946 MeV  in a good agreement with the experimental value 0.929 MeV.
Instead, SLy4, SkM* and SVbas  give much higher energies 1.140 (not shown),
1.462 and 1.622 MeV, respectively. In accordance with the experimental analysis ~\cite{Rob_PRC2008},
all the applied Skyrme forces suggest for the first $2^-$ state in $^{252}$No the
2qp configuration $nn[734\uparrow, 622\uparrow]$. Following Table~\ref{table9:octupoles},
this state demonstrates large $B(E32)$ values.  In the QPM study~\cite{Jol11},
the first $2^-$ state is the lowest among the octupole excitations in $^{252}$No.
We get the same result for SLy6 but not for other forces. Altogether,
SLy6 gives the most relevant description of $2^-$ isomer in $^{252}$No.

One should mention a significant difference in  SLy6 results for the lowest $2^-$ states
in $^{252}$No and $^{254}$No, although in both nuclei these states are dominated  by
the same 2qp configuration $nn[734\uparrow, 622\uparrow]$. Indeed, in $^{252}$No this state has a low
energy 0.95 MeV and large $B(E32)$=11.5 W.u., while in $^{254}$No we have the opposite picture:
high energy 2.12 MeV and low  $B(E32)$=0.6 W.u.. The reason of the difference in $B(E32)$  is that the transition
$734\uparrow \to 622\uparrow$ is particle-hole in $^{252}$No and hole-hole in $^{254}$No (see F-order
in the Table). So this transition in $^{254}$No is strongly suppressed  by the pairing factor
$u_{qq'}=u_q v_{q'} + u_{q'} v_{q}$, where $u_{qq'}$=0.986 in $^{252}$No  and 0.077 in
$^{254}$No. Further, the difference in the excitation energy is caused by different neutron chemical potentials:
$\lambda_n$=-7.09 MeV in $^{252}$No  and -6.35 MeV in $^{254}$No. Note that, in our SLy6 calculations, the neutron
pairing in $^{254}$No is suppressed and its BCS description is rather approximate.

Our results are in accordance  with recent findings for $^{252,254}$No, obtained within
the cranked relativistic  shell-model-like approach ~\cite{Xu_PRL2024}. This study predicts
the large octupole correlations (deformations) $\beta_{32}$ in $^{252}$No and negligible in
$^{254}$No.

Table~\ref{table9:octupoles} shows that, in most of the calculations,  $0^-$ states are more
collective than $1^-$ states.
The calculated $3^-$ states have low excitation energies (1.28-1.48 MeV) and the same (with exception
of SLy6 in  $^{254}$No) dominant 2qp configuration $pp[633\uparrow, 521\downarrow]$.

\subsection{General view}

The above analysis shows that the most relevant description of low-energy spectra
in $^{252,254}$No is provided by the force SLy6.
So, in this section, we limit our discussion to SLy6 (and partly SLy4) results.

In Figures~\ref{fig5:252No_QRPA} and~\ref{fig6:254No_QRPA}, the available experimental data
\cite{nndc} are compared with our SLy6 QRPA results for the lowest excited states with
$K^{\pi} = 0^+, 2^+, 3^+, 4^+, 0^-, 1^-, 2^-, 3^-, 8^-$. The g.s. rotational bands are
demonstrated. For the comparison, SLy4 results for state $K^{\pi} = 0^+, 2^-, 8^-$
are shown.
Figures~\ref{fig5:252No_QRPA} and~\ref{fig6:254No_QRPA} display  a rich spectrum of positive-parity
and negative-parity states above 1 MeV in $^{252,254}$No. Note that energies of some $K> 0$ states
can be 100-300 keV lower if to take into account the pairing blocking effect~\cite{Jol11,Ne_PRC2016}.
The energies of the states in the g.s. rotational band are significantly underestimated, which is explained by too
large value of the calculated moment of inertia (see Table~\ref{No}). So SLy6 obviously underestimates
the pairing in $^{252,254}$No. SLy4 gives a more realistic description of the neutron pairing
in  $^{254}$No and  thus a better agreement for the energy of the first $0^+$ state in this nucleus.
Altogether, SLy6 and SLy4 results are in a reasonable agreement  with available experimental data,

\begin{table}
\caption{The lowest SLy6 neutron and proton 2qp configurations $K=K_1+K_2$ and $K=|K_1-K_2|$
in $^{252,254}$No. The configurations considered previously  are underlined.}
\begin{tabular}{|c|c|c|c|c|}
\hline
$\epsilon_{qq'}$ & qq' & F-order & K$_1$+K$_2$ & K$_1$-K$_2$ \\
\hline
 \multicolumn{5}{|c|}{$^{252}$No} \\
\hline
1.16 &  $pp[521\downarrow, 514\downarrow]$ &F,F+1 & $\underline{4^+}$ & $\underline{3^+}$\\
\hline
1.35  &  $pp[633\uparrow,514\downarrow]$ &F-1,F+1 & 7$^-$ & $\underline{0^-}$ \\
\hline
1.35 &  $pp[633\uparrow, 521\downarrow]$ & F-1,F & 4$^-$ & $\underline{3^-}$\\
\hline
2.06 &  $pp[521\uparrow,521\downarrow]$ & F-2,F & $\underline{2^+}$ & 1$^+$ \\
\hline
2.25 &  $pp[521\uparrow, 633\uparrow]$ &F-2,F-1 & 5$^-$& $\underline{2^-}$ \\
\hline
2.30 &  $pp[633\uparrow, 512\uparrow]$ &F-1,F+3 & 6$^-$ & $\underline{1^-}$ \\
\hline
\hline
1.30 &  $nn[734\uparrow, 622\uparrow]$ & F,F-2 & 7$^-$& $\underline{2^-}$ \\
\hline
1.32 &  $nn[624\downarrow,734\uparrow]$ & F,F+1& $\underline{8^-}$ & $\underline{1^-}$ \\
\hline
2.08  &  $nn[624\downarrow, 743\uparrow]$ & F,F-2 & 7$^-$ & $\underline{0^-}$ \\
\hline
2.33 &  $nn[622\uparrow, 620\uparrow]$ &F-1,F+2 & $\underline{3^+}$ & $\underline{2^+}$\\
\hline
2.34 &  $nn[624\downarrow, 620\uparrow]$ & F,F+2& $\underline{4^+}$ & $3^+$\\
\hline
\hline
\multicolumn{5}{|c|}{$^{254}$No} \\
\hline
1.15 &  $pp[521\downarrow, 514\downarrow]$  & F,F+1& $\underline{4^+}$ & $\underline{3^+}$\\
\hline
1.38 &  $pp[633\uparrow,514\downarrow]$ & F-1,F+1& 7$^-$ & $\underline{0^-}$ \\
\hline
1.38 &  $pp[633\uparrow,521\downarrow]$ &F-1,F & 4$^-$ & $\underline{3^-}$\\
\hline
2.05 &  $pp[521\uparrow,521\downarrow]$ & F-2,F & $\underline{2^+}$ & 1$^+$ \\
\hline
2.27 &  $pp[521\uparrow,633\uparrow]$ &F-2,F-1 & 5$^-$ & $\underline{2^-}$\\
\hline
2.43 &  $pp[633\uparrow,512\uparrow]$ & F-1,F+3& 6$^-$ & $\underline{1^-}$\\
\hline
\hline
1.21 &  $nn[734\uparrow,622\downarrow]$ & F,F+2 & 6$^-$ & $\underline{3^-}$\\
\hline
1.32 &  $nn[622\uparrow, 620\uparrow]$ &F-1,F+1 & $\underline{2^+}$ & 1$^+$ \\
\hline
1.78 &  $nn[734\uparrow,613\uparrow]$ &F,F+3 & 8$^-$ & $\underline{1^-}$\\
\hline
1.89 &  $nn[620\uparrow,613\uparrow]$  &F+1,F+3 & $\underline{4^+}$ & $\underline{3^+}$\\
\hline
2.13 &  $nn[622\uparrow,734\uparrow]$ & F-1,F& 7$^-$ & $\underline{2^-}$\\
\hline
2.17 &  $nn[734\uparrow,615\downarrow]$ & F,F+5& 9$^-$ & $\underline{0^-}$ \\
\hline
\end{tabular}
\label{table10:K12}
\end{table}

Further, Figure~\ref{fig7:B(EL)} shows that, aside the first multipole $K^{\pi}$ states, the energy region
$E <$ 2.5 MeV can contain a lot of next (second, third, etc) QRPA states of a different collectivity.
The most collective are some octupole states with $B(E3)$ of order 10-40 W.u..


The low-energy spectrum can be even richer if we consider positive-parity states with $K > 4$
and negative-parity states with $K>3$, which were omitted in the above analysis.
In Table~\ref{table10:K12}, the lowest proton and neutron 2qp configurations (of particle-hole origin)
for $K^{\pi}=2^+, 3^+, 4^+, 0^-, 1^-, 2^-, 3^-, 8^-$ in $^{252,254}$No are exhibited. Both possible cases
$K=K_1+K_2$ and $K=|K_1-K_2|$ for a given 2qp configuration are considered. The $K^{\pi}$-values realized in
the previously considered excitations are underlined.

Table~\ref{table10:K12} shows that, besides the states considered above,
these configurations can also produce at the same excitation energies negative-parity states
$K^{\pi}=4^-, 5^-, 6^-, 7^-, 9^-$. In $^{252}$No, $7^-$ states
$pp[633\uparrow, 514\downarrow]$ and $nn[734\uparrow, 622\uparrow]$ should coexist at a similar
low energy $\approx$1.3 MeV with $8^-$ isomer $nn[624\downarrow,734\uparrow]$ and $2^-$ isomer
$nn[734\uparrow, 622\uparrow]$.
The low-energy  $3^-$ state  $pp[633\uparrow, 521\downarrow]$ can be accompanied by $4^-$ state
with the similar 2qp structure.
The $3^+$ state  $pp[521\downarrow, 514\downarrow]$ has $4^+$ counterpart.
In $^{254}$No, the low-energy states $0^-$ and $3^-$ have counterparts $7^-$ and $4^-$.
The low-energy $3^+$ and $4^+$ states should have similar energies.

Note that counterparts with $\Delta K=1$ can exhibit Coriolis coupling.
Low-energy 2qp $K^{\pi}=4^-, 5^-, 6^-, 7^-, 4^+$ configurations in $^{254}$No
were also predicted in self-consistent calculations with Gogny force~\cite{De06}.

\section{Conclusions}

A systematic microscopic analysis of low-energy states in even-even isotopes $^{250-262}$No
is performed within a fully self-consistent quasiparticle random-phase approximation (QRPA)
method~\cite{QRPA,Rep_EPJA2017_pairing,Kva_EPJA2019_spurious,Skyax}. A representative set
of Skyrme forces SLy4~\cite{SLy6}, SLy6~\cite{SLy6}, SkM*~\cite{SkM} and SVbas~\cite{SVbas} with
different isoscalar effective masses $m^*/m$ and kinds of pairing (volume and surface)
is used. As compared with SLy6, the force SLy4 avoids the BCS collapse of the neutron pairing in $^{254}$No
and, in this sense, serves to countercheck the SLy6 results.

The low-energy excitations with
$K^{\pi}=0^+, 2^+, 3^+$, $4^+, 0^-, 1^-, 2^-, 3^-, 8^-$ are considered. In addition,
some relevant positive-parity states with $K > 4$ and negative-parity states with $K>3$
are briefly inspected. The main attention is paid to isotopes $^{252}$No and $^{254}$No
which possess the most extensive spectroscopic information on non-rotational states.
Just for these two isotopes a significant shell gap in the neutron s-p
spectrum was earlier predicted~\cite{Rob_PRC2008}.

Among the applied Skyrme forces, the best performance is found for SLy6. This force rather
 well reproduces the energies of the observed $8^-$ and  $2^-$ states in $^{252}$No and
 $3^+$ and $4^+$ states in $^{254}$No. Besides, SLy6 suggests a reasonable
assignment $nn[734\uparrow,613\uparrow]$ for disputed $8^-$ isomer in $^{254}$No.
The similar force SLy4 also demonstrates  a generally good performance. As compared to SLy6,
it provides a much better description of the energy of the observed  $K^{\pi}=0^+$ state in
$^{254}$No. However, SLy4 gives a worse description of the neutron mass staggering quantity
$\delta^{(3)}_{2n}(N,Z)$ and of the energy of $2^-$ isomer in $^{252}$No.

For a more accurate treatment of the pairing decrease arising due to neutron shell gap,
we also apply the forces  UNEDF1~\cite{Kor_UNEDF1}, UNEDF1$^{\rm SO}$~\cite{Shi_PRC2014}
 and UNEDF2~\cite{Kor_UNEDF2}, where the pairing is treated within the Lipkin-Nogami (LN)
 prescription. This prevents the pairing collapse. These forces are implemented for
 alternative description of isomeric states in $^{252}$No and $^{254}$No. However, despite
 a more realistic treatment of the pairing,  the UNEDF forces do not provide a simultaneous
 reasonable description of $8^-, 2^-$ isomers in $^{252}$No and of $3^+, 4^+$ isomers
 in $^{254}$No. It seems that SLy6 suggests a better s-p spectrum and this factor is
 decisive for description of the isomers.

 We assert that just a simultaneous description of well determined  isomers $8^-$,
 $2^-$ in $^{252}$No and of $3^+$  in $^{254}$No can justify the validity of the theory. Instead,
the disputed $8^-$ isomer in $^{254}$No requires very fine tuning of the s-p spectra and pairing and,
in this sense, is not so optimal for testing the theoretical models.

In accordance with the analysis~\cite{Rob_PRC2008}, our SLy6 calculations  predict
in $^{252}$No and, especially in $^{254}$No, a significant shell gap (or, more precisely, a
region of a low level density) in the neutron s-p  spectrum and a corresponding drop of the neutron pairing.
A smaller pairing drop is also obtained in SLy4 and SkM* calculations. The comparison of SLy4
and SLy6 results shows that, in the case of shell gaps,  even tiny changes in the single-particle
spectra can noticeably affect the pairing and related nuclear properties.

The suggested gap in the neutron s-p spectrum can have far-reaching consequences.
First of all, $^{252,254}$No should demonstrate irregularities
in the features of the g.s. rotational bands (moments of inertia, B(E2)-values).
As the next consequence of the pairing drop,
we predict in these nuclei low-energy pairing vibrational  $K^{\pi}= 0^+$ states.
Note that $K^{\pi}= 0^+$ state was observed at 0.888 MeV in $^{254}$No~\cite{Forge_JPCS2023,Forge_PhD}.
This state was treated as a result of coexistence of the normal deformation and superdeformation  or/and
 coexistence of axial and triaxial shapes. Instead, our calculations show that this state should be the
 pairing vibration. Such low-energy $0^+$ pairing vibrations are typical for rare-earth
 and actinide deformed nuclei~\cite{GS_book74,Sol_book76,SSS96,ISS05,IS08}.

 Unlike Fl and Cn isotopes with $Z$ equal or close to the suggested magic number $Z$=114, the nobelium
 isotopes perhaps exhibit a unique case of the suppressed pairing in the mid of the chain of well-deformed
 isotopes whose $Z$ and $N$ are far from the spherical magic numbers. To the best of our knowledge,
  previous studies of $0^+$ states in heavy
 well deformed nuclei did not consider the cases of a suppressed pairing. Low-energy
 $0^+$ pairing vibrations could serve as sensitive indicators of the pairing in superheavy and nearby
 nuclei. Following our study, they can also signal on the shell gaps. It would be interesting to investigate
 these states in photonuclear reactions and various $\gamma$-decay scenarios.

Our calculations show that, in $^{252,254}$No, the lowest  $K^{\pi}=3^+$ and $4^+$ states
are produced by the same 2qp configurations and so should have rather similar energies. This suggestion
is confirmed by the recent observation in $^{254}$No of $4^+$ band head at the energy close to the energy
of the previously known $3^+$ isomer. Because of the similar structure, the $3^+$ and $4^+$ counterparts and their
rotational bands should exhibit a strong Coriolis coupling.

Excitation energies and collectivity of some lowest states, e.g. $K^{\pi}=2^+$, can be very
different in $^{252}$No and $^{254}$No. The calculated octupole states also demonstrate a
variety of excitation energies and 2qp/collective structures. Altogether, a rich multipole
low-energy spectrum is suggested in $^{252}$No and $^{254}$No. In addition to
octupole states, the states  $K^{\pi}=4^-, 7^-$ in $^{252}$No and $4^-, 6^-, 7^-$
in $^{254}$No are predicted.

Note that the applied Skyrme forces are characterized by different isoscalar effective masses
$m^*/m$ and so somewhat deviate in the s-p spectra, shell-gap values and  pairing strength. Thus,
there is a variety in predictions for low-energy 2qp and QRPA states. A further improvement of the
theory is necessary. Nevertheless, even yet unperfect self-consistent methods can
give reasonable interesting predictions.

Further measurements of non-rotational states in $^{252,254}$No are very desirable. Spectroscopy of
 nobelium isotopes could be a crucial test for the
 theory pretending for description of even heavier nuclei, first of all, of superheavy elements.

\section*{ACKNOWLEDGEMENTS}
\label{s5}

V.O.N. and M.A.M. thank Dr. N.N. Arsenyev for useful discussions.
A.R. acknowledges support by the Slovak Research and Development
Agency under Contract No. APVV-20-0532 and by the Slovak grant agency VEGA
(Contract No. 2/0175/24).

\appendix

\section{UNEDF1, UNEDF1$^{\rm SO}$ and UNEDF2 results for $^{252,254}$No.}

 The forces UNEDF1~\cite{Kor_UNEDF1}, UNEDF1$^{\rm SO}$~\cite{Shi_PRC2014}
 and UNEDF2~\cite{Kor_UNEDF2} (called UNEDF hereafter) in the framework of
 Hartree-Fock-Bogoliubov (HFB) and LN methods
are applied  for a more accurate description of $8^-$ isomer  in $^{252,254}$No, where our
 calculations predict a weak neutron pairing. The zero-range pairing interaction (\ref{pairing})
 with $\eta$=0.5 and model parameter $\rho_{\rm pair}$=0.16 ${\rm fm}^{-3}$ is used.
 The calculations are performed with the code \cite{UNEDF_code} using the LN procedure. Note that
 UNEDF1$^{\rm SO}$ was suggested as a spectroscopic-quality force for heavy nuclei~\cite{Shi_PRC2014}.
 As compared with UNEDF1, in this force
 the spin-orbit  and pairing parameters were additionally
 fitted to reproduce the observed spectra of heavy odd nuclei.

 In Table~\ref{mG_UNEDF} we show some force parameters. It is seen
 that UNEDF1$^{\rm SO}$ spin-orbit parameters significantly deviate from
 UNEDF1, UNEDF2 and SLy4-SVbas values (Table~\ref{mG}).

  Table~\ref{UNEDF_pair} shows that, as compared with SLy4-SVbas set,
UNEDF forces give  somewhat smaller equilibrium quadrupole deformations in
$^{252,254}$No. Since LN procedure is used, we avoid the collapse of the
neutron pairing in $^{254}$No. Moreover, here UNEDF1 and UNEDF2
provide the neutron pairing stronger than the proton one.

\begin{table}  
\caption{Isoscalar effective mass $m^*/m$, proton and neutron pairing constants
$G_p$ and $G_n$, IS and IV spin-orbit
parameters $b_4$ and $b'_4$ for UNEDF1,  UNEDF1$^{\rm SO}$, and UNEDF2.}
\begin{tabular}{|c|c|c|c|}
\hline
   & UNEDF1 & UNEDF1$^{\rm SO}$ & UNEDF2 \\
 \hline
   $m^*/m$               & 0.99 & 1.07 & 0.99 \\
   $G_p$ (MeV fm$^3$) & 206.58 & 230.33 & 253.30 \\
   $G_n$ (MeV fm$^3$) & 186.07 & 208.89 & 192.1 \\
   $b_4$ (MeV fm$^5$) &  38.4 & 25.7  & 96.508 \\
   $b'_4$ (MeV fm$^5$)& 71.3 & 77.3 & -16.916 \\
\hline
\end{tabular}
\label{mG_UNEDF}
\end{table}

\begin{table} 
\caption{Quadrupole deformation $\beta$, proton and neutron pairing gaps without
 and with LN addition $\lambda_{2}$ in $^{252,254}$No, calculated
 with  UNEDF1, UNEDF1$^{\rm SO}$ and UNEDF2.}
\begin{ruledtabular}
\begin{tabular}{|l|c|c|c|}
 & UNEDF1   &   UNEDF1$^\mathrm{SO}$  &    UNEDF2     \\
                              \hline
\multicolumn{4}{|c|}{$^{252}$No}  \\
                             \hline
  $\beta$                       & 0.287 & 0.285 & 0.283\\
  $\Delta_p$  (MeV)               & 0.320 & 0.540 & 0.401\\
  $\Delta_p+\lambda_{2p}$ (MeV)   & 0.487 & 0.730 & 0.592\\
  $\Delta_n$ (MeV)                & 0.375 & 0.424 & 0.464\\
  $\Delta_n+\lambda_{2n}$ (MeV)   & 0.561 & 0.608 & 0.672\\

                               \hline
                               \multicolumn{4}{|c|}{$^{254}$No}  \\
                                \hline
  $\beta$                       & 0.285 & 0.282 & 0.280\\
   $\Delta_p$  (MeV)               & 0.318 & 0.534 & 0.402\\
  $\Delta_p+\lambda_{2p}$ (MeV)   & 0.485 & 0.724 & 0.594\\
  $\Delta_n$ (MeV)                & 0.371 & 0.418 & 0.459\\
  $\Delta_n+\lambda_{2n}$ (MeV)   & 0.555 & 0.611 & 0.664\\
 \end{tabular}
\end{ruledtabular}
\label{UNEDF_pair}
\end{table}

\begin{table}  
\caption{The energies (in MeV) of the lowest neutron and proton 2qp $8^-$ states
in $^{252,254}$No, calculated with  UNEDF1, UNEDF1$^{\rm SO}$ and UNEDF2.}
\begin{tabular}{|c|c|c|c|}
\hline
   & UNEDF1 & UNEDF1$^{\rm SO}$ & UNEDF2 \\
\hline
 \multicolumn{4}{|c|}{$^{252}$No, $E_x$= 1.254 MeV}  \\
 \hline
   $nn[624\downarrow,734\uparrow]$ & 0.870 & 0.954 & 1.119\\
   $pp[624\uparrow,514\downarrow]$ & 1.242 & 1.279 & 1.430 \\
 \hline
 \multicolumn{4}{|c|}{$^{254}$No, $E_x$= 1.295 MeV}  \\
 \hline
   $nn[624\downarrow,734\uparrow]$ & 1.036 & 1.141 & 1.338 \\
    $nn[613\uparrow,734\uparrow]$ & 1.081 & 1.321 & 1.605 \\
   $pp[624\uparrow,514\downarrow]$ & 1.190 & 1.250 & 1.390 \\
\hline
\end{tabular}
\label{2qp_UNEDF}
\end{table}

In Table~\ref{2qp_UNEDF}, we show  the calculated  lowest neutron and proton 2qp $K^{\pi}=8^-$
excitations in $^{252,254}$No. In the simple case of $^{252}$No,
only one neutron and one proton 2qp pairs are listed. In a more complicated case of $^{254}$No,
three lowest configurations are demonstrated for each force. The lowest
2qp pair is considered as a priority candidate for the observed $8^-$ isomer.

We see that, in agreement  with our SLy4-SVbas results (Table~\ref{table4:8-}),
UNEDF forces also suggest for $8^-$ isomer in $^{252}$No the neutron configuration
$nn[624\downarrow,734\uparrow]$.  However, for all three UNEDF forces, the energy
of $8^-$ isomer is significantly underestimated. Note that the present UNEDF calculations do not
take into account the pairing blocking effect which should additionally downshift the
calculated isomer energy.

In $^{254}$No, UNEDF forces provide a better description of the isomer energy
than the best  SLy4, SLy6 and SkM* cases shown in Table~\ref{table4:8-}. Instead of
SLy4 and SLy6 assignment $nn[613\uparrow,734\uparrow]$, UNEDF forces
suggest the configuration $nn[624\downarrow,734\uparrow]$.
Note that the energy interval between these two neutron configurations is rather small
(especially for UNEDF1 and  UNEDF1$^{\rm SO}$) and so this UNEDF assignment is rather tentative.
To our opinion, what is indeed important is that UNEDF  forces do not provide a {\it simultaneous}
description of the energies of both $8^-$ isomers in $^{252}$No and $^{254}$No.

In $^{254}$No, the energy intervals between
the lowest neutron and proton 2qp configurations are also small: 164, 109 and 52 keV
for UNEDF1, UNEDF1$^{\rm SO}$ and UNEDF2, respectively.  These values are comparable with
collective shifts due to $\lambda\mu$=98 residual interaction (see discussion in Sec. IV-A). So,
in principle, $8^-$ isomer $^{254}$No can be a mixture of neutron and proton 2qp pairs.

\begin{table}  
\caption{Experimental and calculated energies (in MeV) of $K^{\pi}=2^-$ isomer
$nn[622\uparrow,734\uparrow]$ in $^{252}$No
and $3^+$ isomer  $pp[521\downarrow,514\downarrow]$ in $^{254}$No.}
\begin{tabular}{|c|c|c|c|c|}
\hline
                      & exper & UNEDF1 & UNEDF1$^{\rm SO}$ & UNEDF2 \\
 \hline
   $2^-$ in $^{252}$No & 0.929  & 0.991 & 1.141 & 1.207 \\
   $3^+$ in $^{254}$No & 0.987  & 0.695 & 1.168 & 0.880 \\
\hline
\end{tabular}
\label{2-3+_UNEDF}
\end{table}

Finally, Table~\ref{2-3+_UNEDF} shows that  UNEDF forces do not provide a
simultaneous description of $K^{\pi}=2^-$ isomer $nn[622\uparrow,734\uparrow]$ in $^{252}$No
and $K^{\pi}=3^+$ isomer  $pp[521\downarrow,514\downarrow]$ in $^{254}$No (though perhaps
the UNEDF1$^{\rm SO}$ results can be downshifted to the proper values due to the pairing
blocking effect). At the same time, as shown
in Tables~\ref{table7:3+} and \ref{table9:octupoles},  we get a good description of these isomers
with SLy6. Altogether, one may conclude, that the {\it  simultaneous} description of isomeric states
in $^{252,254}$No  with UNEDF forces is generally not good.  At least, it not better than in
SLy4, SLy6 and SkM* cases. The best general performance is demonstrated by SLy6 force.

\end{document}